\documentclass[aos]{imsart}
\usepackage[utf8]{inputenc}
\usepackage{amsthm}
\usepackage{amssymb}
\usepackage{amsmath}
\usepackage[numbers]{natbib}
\usepackage{graphicx}
\usepackage{float}
\usepackage[T1]{fontenc}
\usepackage{subfigure}
\usepackage{multirow}
\usepackage{appendix}
\usepackage[ruled,linesnumbered]{algorithm2e}
\usepackage{color}
\usepackage{rotating}
\usepackage{adjustbox}
\usepackage{diagbox}
\usepackage{makecell}
\usepackage{caption}
\usepackage{longtable}
\usepackage{graphicx}%插入图片
\usepackage{subfigure} %子图
\usepackage{url}
\usepackage{bm}
\usepackage{algpseudocode}
\usepackage{lastpage}
\usepackage[colorlinks,
	linkcolor=blue,
	anchorcolor=blue,
	citecolor=blue]{hyperref}

\theoremstyle{plain}

\newtheorem{theorem}{Theorem}[section]
\newtheorem{lemma}[theorem]{Lemma}
\newtheorem*{lemma*}{Lemma}

\theoremstyle{remark}

\newtheorem{example}{Example}[section]

\newtheorem{remark}{Remark}[section]

\newtheorem*{assumption*}{Assumption}
\newtheorem*{remark*}{Remark}

\newcommand{\rank}{\operatorname{rank}}
\newcommand{\diag}{\operatorname{diag}}
\newcommand{\tr}{\operatorname{tr}}

\newcommand{\var}{\operatorname*{var}}
\newcommand{\varb}{\mathbb{V}}
\newcommand{\expit}{\operatorname{expit}}

\newcommand{\row}{\operatorname{row}}

\newcommand{\mse}{\operatorname{MSE}}

\newcommand{\aref}[1]{\hyperref[#1]{(#1)}}

\begin{document}
\begin{frontmatter}
\title{Leveraging Shared Factor Structures for Enhanced Matrix Completion with Nonconvex Penalty Regularization}
\runtitle{Shared Factor Structures for Matrix Completion}

\begin{aug}
	\author[A]{\fnms{Yuanhong} \snm{A}\ead[label=e1]{ayh@ruc.edu.cn}},
	\author[A]{\fnms{Xinyan} \snm{Fan}\ead[label=e2]{1031820039@qq.com}},
	\author[B]{\fnms{Bingyi} \snm{Jing}\ead[label=e3]{jingby@sustech.edu.cn}},
	\author[A]{\fnms{Bo} \snm{Zhang*}\ead[label=e4]{mabzhang@ruc.edu.cn}}
	\address[A]{School of Statistics, Renmin University of China \printead[presep={,\\ }]{e1,e2,e4}}
	\address[B]{Department of Statistics and Data Science, Southern University of Science and Technology \printead[presep={,\\ }]{e3}}
\end{aug}

\maketitle

\begin{abstract}%
This article investigates the problem of noisy low-rank matrix completion with a shared factor structure, leveraging the auxiliary information from the missing indicator matrix to enhance prediction accuracy. Despite decades of development in matrix completion, the potential relationship between observed data and missing indicators has largely been overlooked. To address this gap, we propose a joint modeling framework for the observed data and missing indicators within the context of a generalized factor model and derive the asymptotic limit distribution of the estimators. Furthermore, to tackle the rank estimation problem for model specification, we employ matrix nonconvex penalty regularization and establish nonasymptotic probability guarantees for the Oracle property. The theoretical results are validated through extensive simulation studies and real-world data analysis, demonstrating the effectiveness of the proposed method.
\end{abstract}



\end{frontmatter}

\section{Introduction}

Noisy matrix completion, which involves reconstructing a low-rank matrix from partial and noisy observations of its entries, is a problem with wide-ranging applications. These applications include recommendation systems \cite{MCC18, MCL21, MCC22}, causal inference \cite{MCA21, MCP21, xiong2023large}, and portfolio construction in finance \cite{FBI23}. For instance, in a movie recommendation system, users may rate only a few movies, leaving most ratings unobserved. Matrix completion aims to impute these missing entries to provide personalized recommendations based on predicted ratings.

Existing approaches to this problem assume that the distribution of each observed entry $ x_{ij}$ follows $ \mathcal{P}(m_{ij}) $, where $ m_{ij} $ is a parameter belonging to a single-parameter distribution family $ \mathcal{P}(\cdot) $ \cite{MLF22, GFI23, MCL24}. The parameter matrix $\bm{M} = (m_{ij})_{i,j=1}^{n_1,n_2} $ is assumed to be low-rank. Estimating $\bm{M} $ from the partially observed matrix $\bm{X}$ enables the prediction of missing entries as expected values. The estimation methods for $\bm{M} $ can be broadly categorized into two classes: The first category employs regularization techniques such as nuclear norm regularization \cite{ASVT10, MHT10, NOR11, RMC18, TFM23, MCP21, MCP24}, or alternative penalties \cite{HLM11, MNC16, MNC18, NRA13, NNL16, ADM15, MCN20, SSO24}. They are robust as they do not require prior knowledge of the rank of $\bm{M} $. The second category assume that the rank of $\bm{M} $ is known and use matrix factor models \cite{FEM21,MCA21,SIB23,FBI23, xiong2023large, IML24, IMA24}. Their advantage lies in providing the asymptotic distribution of the estimators, enabling inference.

Recently, the modeling of the missing indicator matrix $\bm{W} = (w_{ij})_{i,j=1}^{n_1,n_2} $--where $ w_{ij} = 1 $ indicates that $ x_{ij} $ is observed and $w_{ij} = 0 $ signifies a missing value--has gained significant attention \cite{WNT11, NW12, IPS16, MCC18, MCL21, MCC22, MCC23, MCL24}. These studies primarily assume a Missing at Random (MAR) mechanism, estimate the observation probability $ \pi_{ij} = \mathbb{P}(w_{ij} = 1) $, and use the inverse probability weighting (IPW) method to improve the estimation of $\bm{M} $. This approach has demonstrated improved performance with tighter nonasymptotic upper bounds \cite{MCL21}.

However, the potential relationship between $\bm{X} $ and $\bm{W} $ remains unexplored. For example, in movie rating data, factors such as genre, duration, and production team might simultaneously influence both a user's probability of rating a movie and the actual rating itself. Blockbuster movies, for instance, are more likely to receive ratings and may also achieve higher average ratings. This suggests the existence of common shared factors that influence both $\bm{X} $ and $\bm{W} $. The objective of this work is to model this relationship and improve the estimation of the low-rank matrix $\bm{M} $ using the auxiliary information provided by $\bm{W} $.

To achieve this, we build on the assumption from \cite{MCL21} that the matrix $\bm{W} $ is determined by a low-rank matrix $ \bm{\Theta} = (\theta_{ij})_{i,j = 1}^{n_1,n_2} $, where $ \pi_{ij} = \expit(\theta_{ij}) $. Inspired by the partially shared parameters model \cite{PSP24, LSP22}, we propose a shared factor model in which $ m_{ij} $ and $ \theta_{ij} $ can be decomposed as:
\begin{equation}
    \label{eq:shfactormodel}
    \begin{gathered}
    m_{ij} = \bm{\lambda}_{m1,i}^\top \bm{f}_{s,j} + \bm{\lambda}_{m2,i}^\top \bm{f}_{m,j}, \quad \theta_{ij} = \bm{\lambda}_{\theta1,i}^\top \bm{f}_{s,j} + \bm{\lambda}_{\theta2,i}^\top \bm{f}_{\theta,j},
    \end{gathered}
\end{equation}
where $\bm{\lambda}_{m1,i}, \bm{\lambda}_{\theta1,i}, \bm{f}_{s,j}$ are $d_s \times 1$ vectors, with $\bm{f}_{s,j}$ representing the shared common factor. $\bm{\lambda}_{m2,i}, \bm{f}_{m,j}$ are $d_m$-dimensional vectors corresponding to the specific part of matrix $\bm{M}$, and $\bm{\lambda}_{\theta2,i}, \bm{f}_{\theta,j}$ are $d_{\theta}$-dimensional vectors for the specific part of $\bm{\Theta}$.

There are several empirical evidence that support this model: \cite{MCC18} demonstrated that user attributes such as age and gender significantly influence both the probability of missing data and the observed rating values. These attributes align naturally with the shared factors $\bm{f}_{s,j}$ in our framework. Furthermore, our analysis on the MovieLens 100k dataset \cite{mld16} reveals a high correlation between the linear spaces spanned by the factors of $\bm{M}$ and $\bm{\Theta}$, providing further support for the partially shared structure. By leveraging this structure, we can achieve a 24.02\% improvement in prediction accuracy over existing methods.

In terms of estimation, when the ranks $(d_s, d_m, d_\theta)$ are known, this problem can be framed as a structured matrix factor model and can be solved via maximum likelihood estimation. A key challenge lies in determining these ranks. Since there are three ranks to be determined, the traditional approach: pre-specifying the ranks, estimating the matrix, and then selecting the ranks by minimizing the Information Criteria (IC) \cite{DNF02, IFM03} or cross-validation \cite{FEM21}, become too time-consuming. Furthermore, as demonstrated in \cite{MCN20}, rank estimation based on nuclear regularized estimators tends to overestimate the matrix rank,  potentially leading to an incorrectly specified shared factor model and suboptimal estimation results. To address this problem, we propose a joint estimation framework for $\bm{M}$ and $\bm{\Theta}$ using matrix nonconvex penalty regularization, then estimate these ranks based on the resulting estimators. We demonstrate that, within a broad range of penalty parameters, our method achieves consistent rank estimation. This consistency effectively reduces the problem of selecting three ranks to the simpler task of properly tuning a single penalty parameter.

\subsection{Contributions}

This paper first establishes the limit distribution of factors and factor loadings for various loss functions under the MAR mechanism. This problem is referred to as the generalized factor model, which was initially proposed in \cite{MLF22}, where a single-index factor model was introduced with loss functions ranging from quadratic loss to probit, logit, and other forms to capture the nonlinear relationship between $\bm{X}$ and $\bm{M}$. While this model has been extended to the matrix completion problem under MAR mechanism \cite{GFI23, MCL24}, the limit distribution of the estimator has not yet been established. With known ranks $(d_s,d_m,d_\theta) $, we derive the limit distribution of the proposed estimator and demonstrate that leveraging the shared factor structure reduces the asymptotic mean square error (AMSE). Moreover, we show that this result holds under relaxed moment conditions on the error terms compared to the assumptions in \cite{MLF22}.

Additionally, this paper introduces a framework to analyze the Oracle property--the equivalence of the penalized estimator's behavior to that of the known rank estimator--for nonconvex penalty regularized matrix estimators. While algorithms for matrix estimation with nonconvex penalties have been widely developed over the years \cite{NRA13, ADM15, NNL16, MCN20, SSO24}, and extensive numerical studies suggest the presence of this property, its theoretical guarantees remain largely unexplored. By utilizing the first-order stationary point condition introduced in \cite{SSO24}, we establish a nonasymptotic bound for the Oracle property under the MAR mechanism, providing strong statistical guarantees. It is worth mentioning that our theoretical framework is robust even when the underlying non-penalized part of the optimization problem is nonconvex.

\subsection{Organization of the paper}

The paper is organized as follows: Section \ref{sec:estimationwithknownrank} presents the estimation method under known ranks. Section \ref{sec:estimationwitoutrank} introduces a low-rank matrix estimation approach with nonconvex penalty regularization and a rank estimation method. Section \ref{sec:simulation} discusses simulation results, while Section \ref{sec:realdata} illustrates the proposed method with real data. Finally, Section \ref{sec:conclusion} concludes the paper and outlines future research directions.

\subsection{Notations}

Here, we introduce the notations used throughout this paper. Given an $ n_1 \times n_2 $ matrix $\bm{A} = (a_{ij})_{i,j = 1}^{n_1, n_2} $, we denote the spectral norm, nuclear norm, and Frobenius norm by $ \|\bm{A}\| $, $ \|\bm{A}\|_{\star} $, and $ \|\bm{A}\|_{F} $, respectively. The infinity norm $ \|\bm{A}\|_{\infty} $ is defined as $ \max_{i,j = 1}^{n_1, n_2} |a_{ij}| $, and $ \sigma_{d}(\bm{A}) $ denotes the $ d $-th singular value of $\bm{A} $. For a random variable $X $, we denote $\|X\|_{p} $ as the $l_p$ norm, that $\|X\|_p = (\mathbb{E}[|X|^p])^{1/p} $.

For two sequences of positive random variables $ a_{n} $ and $ b_{n} $, we write $ a_{n} = O_p(b_{n}) $ if, for any $ \epsilon > 0 $, there exists a constant $ C > 0 $ such that $ \limsup_{n \to \infty} \mathbb{P}(a_{n} > C b_{n}) \leq \epsilon $. Similarly, $ a_{n} = o_{p}(b_{n}) $ if $ \lim_{n \to \infty} \mathbb{P}(a_{n} > C b_{n}) = 0 $ for any $ C > 0 $. For positive non-random sequences $ a_{n} $ and $ b_{n} $, the notations $ a_{n} = O(b_{n}) $ and $ a_{n} = o(b_{n}) $ are defined analogously. Additionally, we write $ a_{n} \asymp b_{n} $ if $ a_{n} = O(b_{n}) $ and $ b_{n} = O(a_{n}) $.

In nonasymptotic results, we use $ C $ to denote a constant that may vary from line to line. Following the notation in \cite{weak}, we let $ \|\cdot\|_{\psi} $ denote the Orlicz norm for a convex, non-decreasing function $ \psi: [0, \infty) \to [0, \infty) $ with $ \psi(0) = 0 $, which is widely applied in empirical process theory:
\[
\|X\|_{\psi} = \inf \{C > 0 : \mathbb{E}[\psi(|X|/C)] \leq 1\}.
\]
We define $ \psi_{k} = e^{x^k} - 1 $, which corresponds to the Sub-Exponential and Sub-Gaussian cases when $ k = 1 $ and $ k = 2 $, respectively.

\section{Estimation with Known Ranks}
\label{sec:estimationwithknownrank}

We begin by rewriting the shared factor model from equation \eqref{eq:shfactormodel} into matrix form:

\begin{equation}
\label{eq:shfactormodelmt}
\begin{gathered}
\bm{M} = \bm{\Lambda}_{m1} \bm{F}_s^\top + \bm{\Lambda}_{m2} \bm{F}_m^\top, \\ 
\bm{\Theta} = \bm{\Lambda}_{\theta1} \bm{F}_s^\top + \bm{\Lambda}_{\theta2} \bm{F}_\theta^\top,
\end{gathered}
\end{equation}
where \( \bm{\Lambda}_{m1} \) is the \( n_1 \times d_s \) matrix containing the factor loadings \( \{\bm{\lambda}_{m1,i}\} \), and similarly for the other factor matrices \( \bm{F}_s, \bm{F}_m, \bm{F}_\theta \) and factor loading matrices \( \bm{\Lambda}_{m2}, \bm{\Lambda}_{\theta1}, \bm{\Lambda}_{\theta2} \).

In this context, we employ the symbol \( \cdot \) to denote arbitrary subscripts. For instance, \( \bm{F}_{\cdot} \) may refer to \( \bm{F}_m \), \( \bm{F}_\theta \) or \( \bm{F}_s \), and similarly for \( \bm{\Lambda}_{\cdot} \). To ensure model identification, similar to classical factor models, we impose the following constraints on the matrices \( \bm{\Lambda}_{\cdot} \) and \( \bm{F}_{\cdot} \):
\begin{equation}
\label{eq:restriction}
\begin{gathered}
\bm{\Lambda}_{m1}^\top \bm{\Lambda}_{m1} + \bm{\Lambda}_{\theta 1}^\top \bm{\Lambda}_{\theta 1}, \quad \bm{\Lambda}_{m2}^\top \bm{\Lambda}_{m2}, \quad \bm{\Lambda}_{\theta 2}^\top \bm{\Lambda}_{\theta 2} \text{ are diagonal matrices}, \\ 
\frac{1}{n_2} \bm{F}_{\cdot}^\top \bm{F}_{\cdot} = I, \quad \bm{F}_s^\top \bm{F}_m = \bm{0}_{d_s \times d_m}, \quad \bm{F}_s^\top \bm{F}_\theta = \bm{0}_{d_s \times d_\theta}, \quad \left\| \frac{1}{n_2} \bm{F}_m^\top \bm{F}_\theta \right\| < 1,
\end{gathered}
\end{equation}
where the factors \( \bm{F}_m \) and \( \bm{F}_\theta \) must have no shared column space, which is equivalent to condition \( \left\| \frac{1}{n_2} \bm{F}_m^\top \bm{F}_\theta \right\| < 1 \).

Given the parameter matrices \( \bm{M} \) and \( \bm{\Theta} \), we assume that the observed entries \( x_{ij} \) and the missing indicators \( w_{ij} \) are independent samples from their respective distributions, adhering to the Missing at Random (MAR) assumption. Consequently, the observation probability is given by:
\[
    \Pi_{i,j = 1}^{n_1,n_2} \left[\pi_{ij} \mathbb{P}(x_{ij} | m_{ij})\right]^{w_{ij}} \left( 1 - \pi_{ij} \right)^{1 - w_{ij}},
\]
where \( \pi_{ij} = \expit(\theta_{ij}) \) and \( \mathbb{P}(x_{ij} | m_{ij}) \) represents the probability density function of the distribution \( \mathcal{P}(m_{ij}) \) on \( x_{ij} \).

Since the true density \( \mathbb{P}(x | m) \) may not be accessible, we instead use the negative pseudo-log-likelihood for optimization:

\begin{equation}
\label{eq:loglikelihood}
\mathcal{L}(\bm{M},\bm{\Theta},\eta) = - \sum_{i,j = 1}^{n_1,n_2} \left\{
w_{ij}\left( \log(\pi_{ij}) + \eta l_{ij}\right) + (1 - w_{ij}) \log(1 - \pi_{ij})\right\},
\end{equation}
where \( l_{ij} = l(x_{ij}, m_{ij}) \) represents the pseudo-log-likelihood function for \( x_{ij} \) with parameter \( m_{ij} \), and the parameter \( \eta \) is used to balance the contribution of the information derived from \( x_{ij} \)'s distribution against that from the missing probability part. By adjusting \( \eta \), we can obtain an estimator for \( \bm{M} \) that minimizes the Asymptotic Mean Squared Error (AMSE). Notably, when \( \eta = 1 \) and \( l_{ij} = \log[\mathbb{P}(x_{ij} | m_{ij})] \), the pseudo-log-likelihood function \( \mathcal{L}(\bm{M}, \bm{\Theta}, 1) \) becomes the observed data's negative log-likelihood function with respect to parameters \( \bm{M} \) and \( \bm{\Theta} \).

Minimizing the negative pseudo-log-likelihood function \( \mathcal{L}(\bm{M}, \bm{\Theta}, \eta) \) enables the simultaneous estimation of the parameter matrices \( \bm{M} \) and \( \bm{\Theta} \). When the true ranks of the shared space \( d_s \) and the specific space dimensions \( d_m \) and \( d_\theta \) are known, the shared factor model from equation \eqref{eq:shfactormodelmt} can be estimated by minimizing the function \eqref{eq:loglikelihood}, subject to the constraints in equation \eqref{eq:restriction}. The optimization problem can be formulated as follows:
\begin{equation}
    \label{eq:Oraclepco}
    \begin{gathered}
    \hat{\bm{M}}_o, \hat{\bm{\Theta}}_o = \underset{\bm{M}, \bm{\Theta}}{\arg\min} \mathcal{L}(\bm{M}, \bm{\Theta}, \eta), \\
    \text{subject to } \bm{M} \text{ and } \bm{\Theta} \text{ having the form given by equation \eqref{eq:shfactormodelmt}}, \\
    \text{and } \bm{F}_{\cdot} \text{ and } \bm{\Lambda}_{\cdot} \text{ satisfying the constraints in equation \eqref{eq:restriction}}.
    \end{gathered}
\end{equation}

The iterative gradient descent (IGD) algorithm for solving the optimization problem \eqref{eq:Oraclepco} is shown in Supplement Section 1.
% \ref*{supp-sec:IGD}.

The model assumption specified by \eqref{eq:shfactormodelmt} ensures that the shared factor component \( \bm{F}_s \) can be accurately estimated by incorporating the information from both matrices \( \bm{M} \) and \( \bm{\Theta} \). It is important to note that setting \( d_s = 0 \) in the optimization problem \eqref{eq:Oraclepco} simplifies the model, resulting in separate estimations for \( \bm{M} \) and \( \bm{\Theta} \). Thus, solving the optimization problem in equation \eqref{eq:Oraclepco} allows us to leverage the inherent structure of the data, leading to more accurate estimation of \( \bm{F}_s \) and better imputation of missing values compared to traditional methods.

\begin{remark*}
For the pseudo-log-likelihood function \( l_{ij} = l(x_{ij}, m_{ij}) \), we can choose different forms depending on the data type. Here are some classic examples:
\begin{example}
\label{exa:regression}
\textbf{Regression case:} If \( x_{ij} = m_{ij} + \epsilon_{ij} \), where \( \{\epsilon_{ij}\} \) are independent random variables with mean zero and finite variance, we use:
\[
l_{ij} = l(x_{ij}, m_{ij}) = -\frac{1}{2}(x_{ij} - m_{ij})^{2}.
\]
\end{example}

\begin{example}
\label{exa:glm}
\textbf{Generalized Linear Model (GLM) case:} If \( x_{ij} \) follows a distribution with the probability density function \( \exp\left(\frac{x_{ij} m_{ij} - b(m_{ij})}{\phi} + c(x_{ij}, \phi)\right) \) for some fixed \( \phi \) and known functions \( b(x) \) and \( c(x, \phi) \), we take:
\[
l_{ij} = l(x_{ij}, m_{ij}) = x_{ij} m_{ij} - b(m_{ij}).
\]
\end{example}

\begin{example}
\label{exa:symmetric}
\textbf{Robust Regression case:} If \( x_{ij} \) has a symmetric distribution with median \( m_{ij} \), we can use the Huber loss function:
\[
l_{ij} = l(x_{ij}, m_{ij}) = -
\begin{cases}
\frac{1}{2}(x_{ij} - m_{ij})^{2}, & \text{if } |x_{ij} - m_{ij}| \leq \delta, \\
\delta\left( |x_{ij} - m_{ij}| - \frac{1}{2}\delta \right), & \text{if } |x_{ij} - m_{ij}| > \delta,
\end{cases}
\]
where \( \delta \) is a fixed constant.
\end{example}

\begin{example}
\label{exa:mixture}
\textbf{Mixed-type case:} When the data \( \bm{X} \) contains mixed types, the form of \( l(x_{ij}, m_{ij}) \) can vary according to the data type, as discussed in the literature \cite{MLF22, GFI23, MCL24}.
\end{example}
\end{remark*}

\subsection{Statistical Theory}
In this study, we refer to the true underlying matrices of $\bm{M}$ and $\bm{\Theta}$ as $\bm{M}_{\star}$ and $\bm{\Theta}_{\star}$, respectively. We denote $\pi_{\star,ij} = \expit(\theta_{\star,ij}) $, present the true observed probability. Furthermore, we denote the true factor matrix by $\bm{F}_{\star,\cdot}$ and the true factor loading matrix by $\bm{\Lambda}_{\star,\cdot}$. The following assumptions are made to establish the nonasymptotic bound for the estimator $(\hat{\bm{M}}_{o},\hat{\bm{\Theta}}_{o}) - (\bm{M}_\star,\bm{\Theta}_\star)$.

\begin{assumption*}
\label{as:knowrank}
\item[{(a)}\label{a}] The rank $d_{s}, d_{m}, d_{\theta}$ are fixed, and for given $\bm{M}_{\star}, \bm{\Theta}_{\star}$, $w_{ij}$ are mutually independent with Bernoulli distribution with parameter $\pi_{\star,ij} $, and $x_{ij}$ are mutually independent with distribution $\mathcal{P}(m_{\star,ij})$. Also, $w_{ij}$ are mutually independent with $x_{ij}$.

\item[{(b)}\label{b}] The $\|\cdot\|_{\infty}$ norm of $\bm{\Theta}_{\star}$ and $\bm{M}_{\star}$ are bounded by $\alpha_{\theta}, \alpha_{m}$. And the weight $\eta$ is a fixed number with $0 < \eta < \infty$.

\item[{(c)}\label{c}] For any matrix element $m_{ij} \in [-\alpha_m, \alpha_m]$, there exists $\alpha_{l}$ that:
\[
\pi_{\star,ij}\{\mathbb{E}[l(x_{ij},m_{\star,ij})] - \mathbb{E}[l(x_{ij},m_{ij})]\} \geq \alpha_{l}(m_{ij}- m_{\star,ij})^{2}\text{
for all }\{i,j\}.
\]

Moreover, the function \( l(x, m) \) is differentiable with a uniformly Lipschitz continuous gradient, that we denote \( l'(x, m) = \frac{\partial l(x, m)}{\partial m} \), then there exists a positive constant \( L_l \) such that for any \( m_1, m_2 \in [-\alpha_m, \alpha_m] \):
\[
\max_{i,j = 1}^{n_1,n_2}|l^{\prime}(x_{ij},m_{1}) - l^{\prime}(x_{ij}
,m_{2}) | < L_{l}|m_{1}- m_{2}|.
\]

\item[{(d)}\label{d}] The estimator $(\hat{\bm{M}}_{o},\hat{\bm{\Theta}}_{o})$ of \eqref{eq:Oraclepco} satisfies $\|\hat{\bm{M}}_{o}\|_{\infty}\leq \alpha_{m}$, $\|\hat{\bm{\Theta}}_{o}\|_{\infty} \leq \alpha_{\theta}$. 

For the first order derivative $l'(x_{ij}, m_{\star,ij})$, we use different assumptions:
\begin{itemize}
\item[{(d1)}\label{d1}] Sub-Gaussian: there exists $\alpha_{\psi_2}$ that:
\[
\max_{i,j = 1}^{n_1,n_2}\|l^{\prime}(x_{ij},m_{\star,ij})
\|_{\psi_2}\leq \alpha_{\psi_2}.
\]

\item[{(d2)}\label{d2}] Sub-Exponential: there exists $\alpha_{\psi_1}$ that:
\[
\max_{i,j = 1}^{n_1,n_2}\|l^{\prime}(x_{ij},m_{\star,ij}) \|_{\psi_1}\leq \alpha_{\psi_1}.
\]

The constants satisfy: $\max\{L_{l}\alpha_{m},1/\eta\} \leq \alpha_{\psi_1}(\log (n_{1}n_{2}))$, $\frac{2 L_{l}}{n_{1}^{2}n_{2}^{2}}+ \frac{6 \alpha_{\psi_1}\sqrt{n_1 + n_2}}{(n_{1}n_{2})^{3/2}}\leq \frac{1}{2} \min\{\frac{\exp(\alpha_{\theta})}{2\eta(1 + \exp(\alpha_{\theta}))^{2}} ,\alpha_{l}\}$ and $n_1n_2 \geq e $.

\item[{(d3)}\label{d3}] $8 + \zeta $ moment: there exists $\alpha_{\zeta}> 0$ and $\zeta > 0$ that:
\[
\max_{i,j = 1}^{n_1,n_2}\|l^{\prime}(x_{ij},m_{\star,ij})
\|_{8 + \zeta}\leq \alpha_{\zeta}.
\]

And the constants satisfy: $\alpha_{\zeta}(n_{1}+ n_{2})^{\frac{1}{4 + \zeta / 4}}\geq \max \{2 L_{l}\alpha_{m}, 1/\eta\}$, \\$\frac{L_{l}}{(n_{1}+ n_{2})^{2}} + \frac{ \max\{\alpha_{\zeta},1/\eta\} (8 + \zeta) }{7 + \zeta}\frac{\sqrt{n_{1}n_{2}}}{(n_{1}+ n_{2})^{3/2}}\leq \frac{1}{2}\min\{\frac{\exp(\alpha_{\theta})}{2\eta (1 + \exp(\alpha_{\theta}))^{2}},\alpha_{l}\}$.

\item[{(d4)}\label{d4}] $4 + \delta $ moment: there exists $\alpha_{\delta}> 0$ and $\delta > 0$ that:
\[
\max_{i,j = 1}^{n_1,n_2}\|l^{\prime}(x_{ij},m_{\star,ij})
\|_{4 + \delta}\leq \alpha_{\delta}.
\]

And $\frac{L_{l}}{(n_{1}+ n_{2})^{2}}+ \frac{\max\{\alpha_{\delta},1/\eta\}(4 + \delta)}{3 + \delta}\frac{\sqrt{n_{1}n_{2}}}{(n_{1}+ n_{2})^{3/2}}\leq \frac{1}{2}\min\{\frac{\exp(\alpha_{\theta})}{2\eta (1 +\exp(\alpha_{\theta}))^{2}},\alpha_{l}\}$.

\end{itemize}
\end{assumption*}
Our model is anchored in several fundamental assumptions, each of which is firmly supported by existing literature. Assumption \aref{a} pertains to the low-rank structure of matrices, employing a logistic regression model for the missing indicator matrix as described in \cite{MCL21}, and a generalized factor model for the data matrix as outlined in \cite{MLF22}. Assumption \aref{b} is a standard presupposition ensuring that model parameters are maintained within a viable range. Assumption \aref{c} is the property of function $l(x,m)$, which is divided into two parts: the expectation lower bound assumption is quite wild; for example, the cases presented in Examples \ref{exa:regression} and \ref{exa:glm} for the convex function $b(m)$ meet this criterion. This assumption also corresponds to assumption (ii) in \cite{QFM21}. The Lipschitz assumption, is commonly utilized in optimization theory, as seen in \cite{NNL16, SSO24}. It is important to note that we do not require the function $l(x,m)$ to be convex with respect to $m$, which means the function $\mathcal{L}(\bm{M},\bm{\Theta},\eta) $ doesn't need to be convex. 
Here we don't need the Restrict Strong Convex property \cite{negahbanUnifiedFrameworkHighdimensional2012,NW12} to build the estimator error bound, allowing our method to handle more complex loss function.
 Assumption \aref{d} encompasses three components: The constraint on the infinity norm of the estimators $\hat{\bm{M}}_{o}$ and $\hat{\bm{\Theta}}_{o}$ is primarily included to ensure theoretical guarantees. By assigning sufficiently large values to the constants $\alpha_{m}$ and $\alpha_{\theta}$, this constraint becomes feasible; For the inequality constraint of constants $\alpha_\cdot $ and $\eta $ in {\aref{d2} $\sim $ \aref{d4}}, as we can take $n_1,n_2 $ large enough, the assumption is easily satisfied; Furthermore, the tail bound assumptions on $l'(x_{ij}, m_{\star,ij})$ are essential for managing the behavior of the estimators' error terms.

For the sake of notation, we denote \(\bm{H} = (\bm{M}^{\top}, \bm{\Theta}^{\top})^{\top} \) as the total parameter matrix, with its estimator $\hat{\bm{H}}_o = (\hat{\bm{M}}_o^\top, \hat{\bm{\Theta}}_o^\top)^\top $. Then from the representation \eqref{eq:shfactormodelmt}, the ranks of \( \bm{M} \), \( \bm{\Theta} \), and \(\bm{H} \) are \( d_s + d_m \), \( d_s + d_\theta \) and \( d = d_s + d_m + d_\theta \), respectively. The nonasymptotic probability tail bond of $\hat{\bm{H}}_o - \bm{H}_\star $ is given in the following theorem.

\begin{theorem}
\label{th:Oracleproperty}
Under Assumptions {\aref{a} $\sim$ \aref{c}} and different \aref{d}, there exist universal constants $C_{1},C_{2}$:
\begin{itemize}
\item under Sub-Gaussian Assumption \aref{d1}:
\[
\mathbb{P}(\|\hat{\bm{H}}_{o}- \bm{H}_{\star}\|_{F}\geq t) \leq C_{1}\exp (
- C_{2}\frac{\alpha_{h}^{2}}{c_{\psi_2}^{2}d}\frac{t^{2}}{n_{1}+
n_{2}}),
\]
where $\alpha_{h}= \min\{\frac{\exp(\alpha_{\theta})}{2(1 + \exp(\alpha_{\theta}))^{2}} , \eta \alpha_{l}\}$ and $c_{\psi_2}= \max\{1/\sqrt{\log(2)}, 2\eta /\sqrt{\log(2)} L_{l}\alpha_{m}+ \eta \alpha_{\psi_2}\}$.

\item under Sub-Exponential Assumption \aref{d2}, when $t \geq \frac{\log(n_{1}n_{2})}{\sqrt{n_{1}n_{2}}}$:
\[
\mathbb{P}(\|\hat{\bm{H}}_{o}- \bm{H}_{\star}\|_{F}\geq t) \leq \frac{2}{n_{1}n_{2}}
+ C_{1}\exp(-C_{2}\frac{\alpha_{h}^{2}}{\eta^{2}\alpha_{\psi_1}^{2}d}
\frac{t^{2}}{(n_{1}+ n_{2})(\log(n_{1}n_{2}))^{2}}).
\]

\item under $8 + \zeta$ moment Assumption \aref{d3}, when $t \geq (n_{1}+ n_{2})^{-\frac{8+ 5 \zeta}{32 + 2 \zeta}}$:
\[
\mathbb{P}(\|\hat{\bm{H}}_{o}- \bm{H}_{\star}\|_{F}\geq t) \leq \frac{n_{1}n_{2}}{(n_{1}+
n_{2})^{2 + \frac{2\zeta}{16 + \zeta}}}+ C_{1}\exp (- C_{2}\frac{\alpha_{h}^{2}}{
\eta^{2}\alpha_{\zeta}^{2}d}\frac{t^{2}}{(n_{1}+ n_{2})^{1 + 2/(4
+ \zeta / 4)}}).
\]

\item under $4 + \delta$ moment Assumption \aref{d4}, when $t > (n_{1}+ n_{2})^{- \delta / 2}$:
\[
\mathbb{P}(\|\hat{\bm{H}}_{o}- \bm{H}_{\star}\|_{F}\geq t ) \leq \frac{n_{1}n_{2}}{(n_{1}+
n_{2})^{2 + \delta/2}}+ C_{1}\exp(- C_{2}\frac{\alpha_{h}}{c_{\delta}(\alpha_{m}+
\alpha_{\theta}) d^{3/2}}\frac{t^{2}}{(n_{1}+ n_{2})^{3/2}}),
\]
where $c_{\delta}= \max\{\eta \alpha_{\delta}+ 2 \eta L_{l}\alpha_{m},1\}$. 
\end{itemize}
\end{theorem}

In Theorem \ref{th:Oracleproperty}, we employ the classic chaining and peeling technique with matrix Frobenius norm, as demonstrated in \cite{weak}, to construct the nonasymptotic bounds under Assumptions \aref{d1}. Additionally, we utilize the truncation method to achieve results under Assumptions \aref{d2} and \aref{d3}. For Assumption \aref{d4}, we replace the matrix Frobenius norm with the matrix infinity norm in chaining procedural to derive the error bound.

It is noteworthy that under Assumptions \aref{d1}, \aref{d2}, \aref{d3} and \aref{d4}, our theorem indicates that the estimator's error term $\|\hat{\bm{H}}_{o}- \bm{H}_{\star}\|_{F}$ is of the order $O_p(\sqrt{n_{1}+ n_{2}})$, $O_p(\log(n_{1}n_{2}) \sqrt{n_{1}+ n_{2}})$, $o_{p}((n_{1}+ n_{2})^{3/4})$ and $O_p((n_{1}+ n_{2})^{3/4})$, respectively. Furthermore, for Assumption \aref{d4}, we can show the following Lemma:

\begin{lemma}
\label{le:Oraclepropertyd4}
Under Assumptions {\aref{a} $\sim$ \aref{c}} 	and \aref{d4}, there exist universal constants $C_{1}, C_{2}$ that for any fixed $c > 0$ and $t > (n_{1}+ n_{2})^{-\delta/2}$, we have:
\begin{equation*}
\begin{gathered}
\mathbb{P}(\|\hat{\bm{H}}_{o} - \bm{H}_\star\|_F \geq t) \leq \mathbb{P}(\frac{\|\hat{\bm{H}}_{o}-
\bm{H}_{\star}\|_{F}}{(n_{1}+ n_{2}) \|\hat{\bm{H}}_{o}- \bm{H}_{\star}\|_{\infty}} > c)+
\\ \frac{n_{1}n_{2}}{(n_{1}+ n_{2})^{2 + \delta/2}} + C_1 \exp(-C_2
\frac{\alpha_{h}c }{c_{\delta}d^{3/2}(n_{1}+ n_{2})^{1/2}} t).
\end{gathered}
\end{equation*}

Moreover, when $t \geq 1$, we have:
\begin{gather*}
\mathbb{P}(\frac{\|\hat{\bm{H}}_{o}- \bm{H}_{\star}\|_{F}^{2 + c}}{(n_{1}+ n_{2})^{1
+ c}\|\hat{\bm{H}}_o - \bm{H}_{\star}\|_{\infty}^{1 + c}}\geq t) \leq \frac{n_{1}n_{2}}{(n_{1}+n_{2})^{2 + \delta/2}}+ \\ C_{1}(c + \frac{1 }{c})\exp(-C_{2}\frac{\alpha_{h}}{c_{\delta}d^{3/2}(n_{1}+
n_{2})^{1/2}}(t/2)^{2/(2+c)}).
\end{gather*}
\end{lemma}

For the $4+\delta$ moment derivative Assumption \aref{d4}, as we have shown in Lemma \ref{le:Oraclepropertyd4}, we derive the error bound $\|\hat{\bm{H}}_{o}- \bm{H}_{\star}\|_{F}= O_p((n_{1}+ n_{2})^{3/2}\|\hat{\bm{H}}_{o}- \bm{H}_{\star}\|_{\infty}/ \|\hat{\bm{H}}_{o} - \bm{H}_{\star}\|_{F})$. These findings represent a novel contribution to the existing literature. The derived error bounds are crucial for understanding the asymptotic behavior of the estimator, as we will show in the following sections.

Theorem \ref{th:Oracleproperty} delineates the behavior of the total error term. Now we turn to show the error bound for the shared and individual components of the matrix $\hat{\bm{M}}_o $ and $\hat{\bm{\Theta}}_o$. We first introduce additional notations.

Let the shared component of $\bm{M}_{\star}$ be denoted as $\bm{M}_{\star,sh}= \bm{\Lambda}_{\star,m1}^{\top}\bm{F}_{\star,s}$, and its individual component as $\bm{M}_{\star,in}= \bm{\Lambda}_{\star,m2}^{\top}\bm{F}_{\star,m}$. We estimate these components with $\hat{\bm{M}}_{o,sh}= \hat{\bm{\Lambda}}_{o,m1}^{\top}\hat{\bm{F}}_{o,s}$ and $\hat{\bm{M}}_{o,in}= \hat{\bm{\Lambda}}_{o,m2}^{\top}\hat{\bm{F}}_{o,m}$, respectively. The same notation applies to the definitions of $\bm{\Theta}_{\star,sh}$, $\bm{\Theta}_{\star,in}$, $\hat{\bm{\Theta}}_{o,sh}$, and $\hat{\bm{\Theta}}_{o,in}$. Then, with additional assumptions, we can derive the error bound for shared and individual components:

\begin{assumption*}
\begin{itemize}
\item[{(e)}\label{e}] For matrix $\bm{M}_{\star}$ and $\bm{\Theta}_{\star}$:
\begin{align*}
\quad \frac{1}{\sigma_{1}(\bm{M}_{\star})}\min_{1 \leq i \leq d_{s} + d_m}\{\sigma_{i}(\bm{M}_{\star}) - \sigma_{i+1}(\bm{M}_{\star})\} \geq \kappa_{d}> 0, \quad \frac{\sigma_{1}(\bm{M}_{\star})}{\sigma_{d_{s} + d_m}(\bm{M}_{\star})}\leq \kappa_{m}< \infty,         \\
\frac{1}{\sigma_{1}(\bm{\Theta}_{\star})}\min_{1 \leq i \leq d_{s} + d_\theta}\{\sigma_{i}(\bm{\Theta}_{\star}) - \sigma_{i+1}(\bm{\Theta}_{\star})\} \geq \kappa_{d}> 0,\quad \frac{\sigma_{1}(\bm{\Theta}_{\star})}{\sigma_{d_{s} + d_\theta}(\bm{\Theta}_{\star})}\leq \kappa_{m}< \infty.
\end{align*}
\end{itemize}
\end{assumption*}
We utilize the parameter $\kappa_{d}$ to regulate the variability among distinct singular values of the matrix. Additionally, $\kappa_{m}$ is employed to control the disparity between the maximum and minimum non-zero singular values. Drawing from the matrix's eigenvectors' perturbation theory,  we can effectively manage the difference between the singular eigenvectors of $\hat{\bm{M}}_o$ and $\bm{M}_{\star}$. This control is achieved through the parameters $\kappa_{d}$, $\kappa_{m}$, and the norm $\|\hat{\bm{M}}_o - \bm{M}_{\star}\|$. Similar assumptions are applied to $\bm{\Theta}_{\star}$.

We denote $c_{m,\theta}= 20 \frac{\kappa_{m}}{\kappa_{d}}(\frac{\sqrt{2d_{s}+ 2d_{m}}}{\|\bm{M}_{\star}\|} + \frac{\sqrt{2d_{s}+ 2d_{\theta}}}{\|\bm{\Theta}_{\star}\|})$, and define the bound $\xi = \|\frac{1}{n_{2}}\bm{F}_{\star,m}^{\top}\bm{F}_{\star,\theta}\|$ that $\xi < 1 $. When the estimated error $\|\hat{\bm{H}}_{o}- \bm{H}_{\star}\|_{F}$ isn't excessively large, we can control the error of shared and individual components:

\begin{theorem}
\label{th:shincom}
When $\|\hat{\bm{H}}_{o}- \bm{H}_{\star}\| \leq \frac{1 - \xi}{8 c_{m,\theta}}$, we have the bound:
\[
\max\{\|\hat{\bm{M}}_{o,sh}-\! \bm{M}_{\star,sh}\|,\! \|\hat{\bm{M}}_{o,in}-\! \bm{M}_{\star,in}\|\} \!
\leq \! 2\|\hat{\bm{H}}_{o} \! - \! \bm{H}_{\star}\| \! +\! \|\bm{M}_{\star}\|(\frac{5}{1 \! - \! \xi^2} \! + \! 1)c_{m,\theta}
\|\hat{\bm{H}}_{o} \! - \! \bm{H}_{\star}\|,
\]
\[
\max\{\|\hat{\bm{\Theta}}_{o,sh}-\! \bm{\Theta}_{\star,sh}\|,\! \|\hat{\bm{\Theta}}_{o,in} \! - \!
\bm{\Theta}_{\star,in}\|\} \! \leq \! 2\|\hat{\bm{H}}_{o} \! - \! \bm{H}_{\star}\| \! + \! \|\bm{\Theta}_{\star}
\|(\frac{5}{1 \! - \! \xi^2} \! + \! 1)c_{m,\theta}\|\hat{\bm{H}}_{o} \! - \! \bm{H}_{\star}\|.
\]
\end{theorem}
It is important to note that when $\|\bm{M}_{\star}\|$ and $\|\bm{\Theta}_{\star}\|$ are of the order $\sqrt{n_{1}n_{2}}$, and the ratio $\frac{\kappa_{m}}{\kappa_{d}}$ is of constant order, then $c_{m,\theta}$ is of the order $(n_{1}n_{2})^{-1/2}$. As demonstrated in Theorem \ref{th:Oracleproperty}, the error of the total parameter matrix $\|\hat{\bm{H}}_{o} - \bm{H}_{\star}\|$ is of the order $(n_1 + n_2)^{3/4}$. Consequently, the condition $ \|\hat{\bm{H}}_{o} - \bm{H}_{\star}\| \leq \frac{1 - \xi}{8 c_{m,\theta}} $ is satisfied for sufficiently large $n_{1}$ and $n_{2}$ with high probability when $n_1 \asymp n_2 $. Theorem \ref{th:shincom} further illustrates that the error in the shared and individual components is of the same order as $\|\hat{\bm{H}}_{o} - \bm{H}_{\star}\|$.

Our investigation shifts focus towards the limit distribution of the factor's estimator $\hat{\bm{F}}_{o,\cdot}$ and the factor loading's estimator $\hat{\bm{\Lambda}}_{o,\cdot}$. We can establish the central limit theorem with the support of additional assumptions.

\begin{assumption*}

\item[{(f)}\label{f}] This assumption is divided into five parts:
\begin{itemize}
\item[{(f1)}\label{f1}] For the factor loading matrix $\bm{\Lambda}_{\star,\cdot}$ and factor matrix $\bm{F}_{\star,\cdot}$ satisfy:
\begin{gather*}
\frac{1}{n_{2}}\bm{F}_{\star,\cdot}^{\top}\bm{F}_{\star,\cdot}= \bm{I}_{d \cdot}
 \text{ are identical matrices},\\
  \frac{1}{n_{1}}\bm{\Lambda}_{\star,m1}^{\top}\bm{\Lambda}_{\star,m1}+
\bm{\Lambda}_{\star,\theta 1}^{\top}\bm{\Lambda}_{\star,\theta 1}, \ \frac{1 }{n_1 }  \bm{\Lambda}_{\star,m2}^{\top}\bm{\Lambda}_{\star,m2} \text{ and } \frac{1}{n_1} \bm{\Lambda}_{\star,\theta 2}^{\top}\bm{\Lambda}_{\star,\theta 2} \text{ are diagonal matrix }\\ 
\text{denoted as } d_s \times d_s \text{ matrix } \bm{D}_{n_1,s} ,\  d_m \times d_m \text{ matrix } \bm{D}_{n_1,m} \text{ and } d_\theta \times d_\theta \text{ matrix } \bm{D}_{n_1,\theta}, \\ \lim_{n_1,n_2
\to \infty}\bm{D}_{n_1,\cdot}= \bm{D}_{\cdot}\ \text{with different
diagonal elements $\bm{D}_{\cdot,i}> \bm{D}_{\cdot,i+1}> 0$}.
\end{gather*}

\item[{(f2)}\label{f2}] The row and column dimensions grow simultaneously  that $n_{1}\asymp n_{2}$. 

\item[{(f3)}\label{f3}] The function $l(x,m)$ is second order continuous differential on $m_{ij} \in [- \alpha_m, \alpha_m] $, with $\sup_{-\alpha_m \leq m \leq \alpha_m} |l''(x_{ij},m)| \leq L_{l}$. And the second order derivative $l^{\prime\prime}(x_{ij},m) $ is Lipschtiz continuous with Lipschtiz constant $L_l $ for $m \in [- \alpha_m, \alpha_m] $.

\item[{(f4)}\label{f4}] The parameters $L_l, \alpha_{\cdot}$ in Assumption \aref{b} $\sim$ \aref{d} and $\xi $ are constant that uncorrelated with $n_{1}, n_{2}$. 

\item[{(f5)}\label{f5}] The estimator error $\|\hat{\bm{H}}_o - \bm{H}_\star\|_F $ has the following order:
\[
\| \hat{\bm{H}}_{o}- \bm{H}_{\star}\|_{F} = o_{p}((n_{1}+ n_{2}
)^{3/4}).
\]
\end{itemize}
\end{assumption*}

Assumption \aref{f1} embodies a strong factor model assumption that incorporates a shared factor structure, crucial for parameter identification as demonstrated in the literature \cite{IFM03}. Assumption \aref{f2} is a commonly used assumption for the factor model, as seen in \cite{QFM21}. Assumption \aref{f3} is a standard assumption for the function $l(x,m)$ to control the Taylor expansion residual term, and the common functions, such as Example \ref{exa:regression} and \ref{exa:glm} when $b(m) $ is third order continuous differential, satisfy this assumption. Assumption \aref{f4} is a standard assumption, and by selecting sufficiently large or small values for the constants $L_l, \alpha_\cdot, \xi$, this assumption is rendered feasible. Assumption \aref{f5} is a pivotal assumption to control the order of the Taylor expansion's remainder.

Regarding the nonasymptotic behavior of the estimator error term $\hat{\bm{H}}_{o} - \bm{H}_{\star}$, our analysis presented in Theorem \ref{th:Oracleproperty} reveals that Assumption \aref{f5} holds under Assumptions \aref{d1}, \aref{d2}, and \aref{d3}. For \aref{d4}, with the aid of Lemma \ref{le:Oraclepropertyd4},  the Assumption \aref{f5} is satisfied with an additional uniformity property of $\hat{\bm{H}}_{o} - \bm{H}_{\star}$, which is expressed as:
\begin{equation}
\label{eq:uniformityassumption}
\frac{\sqrt{n_1 n_2}\|\hat{\bm{H}}_{o}- \bm{H}_{\star}\|_{\infty}}{\|\hat{\bm{H}}_{o}- \bm{H}_{\star}\|_{F}}= o_{p}((n_{1}+ n_{2})^{1/4}).
\end{equation}

For the uniformity property, the term $\frac{\|\hat{\bm{H}}_{o} - \bm{H}_\star\|_{F}}{\sqrt{n_{1}n_{2}}}$ represents the square root of the average square error. Assumption \eqref{eq:uniformityassumption} stipulates that the maximum error of $\hat{\bm{H}}_{o} - \bm{H}_\star$ should not exceed the average error by a significant margin, with a specific upper bound of $(n_{1}+ n_{2})^{1/4}$.

As shown in model \eqref{eq:restriction} with Assumption \aref{f1}, we can take diagonal matrix $\bm{D}_{1},\bm{D}_{2},\bm{D}_{3}$ with diagonal element equal $1$ or $-1$ , that:
\[
\bm{D}_{1}= \diag(\underbrace{\pm 1, \cdots, \pm 1}_{d_s}), \quad \bm{D}_{2}= \diag(\underbrace{\pm
1, \cdots, \pm 1}_{d_{\theta}}), \quad \bm{D}_{3}= \diag(\underbrace{\pm 1, \cdots,
\pm 1}_{d_m}),
\]
where $\bm{D}_{1}$ is the diagonal $\pm 1$ matrix minimize $\|\hat{\bm{\Lambda}}_{o,m1}\bm{D}_{1}- \bm{\Lambda}_{\star,m1}\|_{F}+ \|\hat{\bm{\Lambda}}_{o, \theta 1}\bm{D}_{1}- \bm{\Lambda}_{\star, \theta 1}\|_{F}$, and $\bm{D}_{2}, \bm{D}_{3}$ are defined similarly to minimize $\|\hat{\bm{\Lambda}}_{o,\theta 2}\bm{D}_{2}- \bm{\Lambda}_{\star, \theta 2}\|_{F}$ and $\|\hat{\bm{\Lambda}}_{o, m2} \bm{D}_{3} - \bm{\Lambda}_{\star, m2}\|_{F}$. Then we have the following central limit theorem.
\begin{theorem}
\label{th:qcoOracle}
With the Assumption {$\aref{a} \sim \aref{f}$}, we have the central limit theory of $\hat{\bm{\Lambda}}_{o,m\cdot}, \hat{\bm{\Lambda}}_{o,\theta\cdot} , \hat{\bm{F}}_{o,\cdot}$, and also the estimator for $\hat{m}_{o,ij}$ and $\hat{\theta}_{o,ij}$:

\begin{gather*}
\sqrt{n_{2}}\bm{\Phi}_{\theta,i}^{1/2}(
\begin{pmatrix}
\bm{D}_{1}\hat{\bm{\lambda}}_{o,\theta 1,i} \\
\bm{D}_{2}\hat{\bm{\lambda}}_{o,\theta 2,i}
\end{pmatrix}
-
\begin{pmatrix}
\bm{\lambda}_{\star,\theta 1,i} \\
\bm{\lambda}_{\star,\theta 2,i}
\end{pmatrix}) \to \mathcal{N}(0, \bm{I}_{d_s + d_\theta}), \\ \sqrt{n_{2}}\bm{\Phi}
_{m,i}(\tilde{\bm{\Phi}}_{m,i})^{-1/2}(
\begin{pmatrix}
\bm{D}_{1}\hat{\bm{\lambda}}_{o,m1,i} \\
\bm{D}_{3}\hat{\bm{\lambda}}_{o,m2,i}
\end{pmatrix}
-
\begin{pmatrix}
\bm{\lambda}_{\star,m1,i} \\
\bm{\lambda}_{\star,m2,i}
\end{pmatrix}) \to \mathcal{N}(0, \bm{I}_{d_s + d_m}), \\ \sqrt{n_{1}}\bm{\Psi}_{i}
(\tilde{\bm{\Psi}}_{i})^{-1/2}(
\begin{pmatrix}
\bm{D}_{1}\hat{\bm{f}}_{o,s,i}     \\
\bm{D}_{2}\hat{\bm{f}}_{o,\theta,i} \\
\bm{D}_{3}\hat{\bm{f}}_{o,m,i}
\end{pmatrix}
-
\begin{pmatrix}
\bm{f}_{\star,s,i}     \\
\bm{f}_{\star,\theta,i} \\
\bm{f}_{\star,m,i}
\end{pmatrix}) \to \mathcal{N}(0, \bm{I}_{d_s + d_m + d_\theta}),\\ \sqrt{n_{1}+
n_{2}}\varb_{\theta,ij}^{-1/2}(\hat{\theta}_{o,ij}- \theta_{\star,ij}) \to
\mathcal{N}(0,1), \\ \sqrt{n_{1}+ n_{2}}\varb_{m,ij}^{-1/2}(\hat{m}_{o,ij}
- m_{\star,ij}) \to \mathcal{N}(0,1), \\ \varb_{\theta,ij}= \frac{n_{1}+
n_{2}}{n_{2}}
\begin{pmatrix}
\bm{f}_{\star,s,j}     \\
\bm{f}_{\star,\theta,j}
\end{pmatrix}^{\top}\bm{\Phi}_{\theta,i}^{-1}
\begin{pmatrix}
\bm{f}_{\star,s,j}     \\
\bm{f}_{\star,\theta,j}
\end{pmatrix}
+ \frac{n_{1}+ n_{2}}{n_{1}}
\begin{pmatrix}
\bm{\lambda}_{\star,\theta 1,i}  \\
\bm{\lambda}_{\star,\theta 2, i} \\
\bm{0}
\end{pmatrix}^{\top}\bm{\Psi}_{j}^{-1}\tilde{\bm{\Psi}}_{j}\bm{\Psi}_{j}^{-1}
\begin{pmatrix}
\bm{\lambda}_{\star,\theta 1,i}  \\
\bm{\lambda}_{\star,\theta 2, i} \\
\bm{0}
\end{pmatrix}, \\ \varb_{m,ij}= \frac{n_{1}+ n_{2}}{n_{2}}
\begin{pmatrix}
\bm{f}_{\star,s,j} \\
\bm{f}_{\star,m,j}
\end{pmatrix}^{\top}\bm{\Phi}_{m,i}^{-1}\tilde{\bm{\Phi}}_{m,i}\bm{\Phi}_{m,i}^{-1}
\begin{pmatrix}
\bm{f}_{\star,s,j} \\
\bm{f}_{\star,m,j}
\end{pmatrix}
+ \\ 
\frac{n_{1}+n_{2}}{n_{1}}
\begin{pmatrix}
\bm{\lambda}_{\star,m1,i} \\
\bm{0}\\
\bm{\lambda}_{\star,m2,i}
\end{pmatrix}^{\top}\bm{\Psi}_{j}^{-1}\tilde{\bm{\Psi}}_{j}\bm{\Psi}_{j}^{-1}
\begin{pmatrix}
\bm{\lambda}_{\star,m1,i} \\
\bm{0}\\
\bm{\lambda}_{\star,m2,i}
\end{pmatrix},
\end{gather*}
where $\bm{\Phi}_{\cdot,i}, \tilde{\bm{\Phi}}_{m,i}, \bm{\Psi}_{j}, \tilde{\bm{\Phi}}_{j}$ are defined in Appendix \ref{sec:notation}, and $\bm{\lambda}_{\cdot,i}$, $\bm{f}_{\cdot,i}$ represent the transportation of the $i$-th row of $\bm{\Lambda}_{\cdot}$ and $\bm{F}_{\cdot}$ respectively.
\end{theorem}

Theorem \ref{th:qcoOracle} delineates the limit distribution of the estimators, thereby enabling the construction of confidence intervals for these estimators and facilitating a comparison of the AMSE with other methods. It is important to note that the limit distribution of $\hat{\bm{\lambda}}_{o,\cdot}$ is equivalent to that which would be obtained by optimizing the function $\mathcal{L}(\bm{M},\bm{\Theta},\eta)$ on $\bm{F}_{\star,\cdot}$, effectively treating the factors as if they were observable. This principle is similarly applicable to the limit distribution of $\hat{\bm{f}}_{o,\cdot}$.

This theorem is established by utilizing the factor and factor loading matrices parameter's Taylor expansion technique as proposed by \cite{MLF22,IML24}, in conjunction with the Burkholder-Davis-Gundy inequality. Unlike the $(14 + \delta)$-th moment assumption required by the approach in \cite{MLF22} and the binary distribution assumption in \cite{SIB23}, our method necessitates only an $(8 + \delta)$-th moment assumption for some $\delta > 0$. Alternatively, with a $(4 + \delta)$-th moment assumption coupled with the uniformity assumption \eqref{eq:uniformityassumption}, we can achieve the same result.

To make inferences on \( \hat{\bm{\lambda}}_{o,\star} \) and \( \hat{\bm{f}}_{o,\star} \), one requires consistent estimators for \( \bm{\Phi}_{\cdot,i}, \tilde{\bm{\Phi}}_{m,i}, \bm{\Psi}_i \), and \( \tilde{\bm{\Psi}}_i \). By replacing the expectations $\mathbb{E}[l''(x_{ij}, m_{\star,ij})], \mathbb{E}[|l'(x_{ij}, m_{\star,ij})|^2]  $ with the sample $l''(x_{ij}, \hat{m}_{o,ij}) $ and $|l'(x_{ij}, \hat{m}_{o,ij})|^2 $, the real parameter $\pi_{\star,ij} $, $ \theta_{\star,ij}$, $\bm{f}_{\star,\cdot}$, $\bm{\lambda}_{\star,\cdot}$ with estimator $\expit(\hat{\theta}_{o,ij}) $, $ \hat{\theta}_{o,ij}$, $\hat{\bm{f}}_{o,\cdot}$, $\hat{\bm{\lambda}}_{o,\cdot}$, respectively, in the formulas provided in Appendix \ref{sec:notation}, we obtain the estimators \( \hat{\bm{\Phi}}_{\cdot,i}, \hat{\tilde{\bm{\Phi}}}_{m,i}, \hat{\bm{\Psi}}_i \), and \( \hat{\tilde{\bm{\Psi}}}_i \).

Additionally, the variances \( \varb_{\theta,ij} \) and \( \varb_{m,ij} \) can be estimated by substituting the matrices and vectors in their definitions with their corresponding estimators. This yields \( \hat{\varb}_{\theta,ij} \) and \( \hat{\varb}_{m,ij} \) as the estimators for the variances. The proof of these estimators' consistency is provided in the Supplement Section 12.
% \ref*{supp-sec:consistency}.

As the matrices $\bm{\Psi}_j, \tilde{\bm{\Psi}}_j$ are functions of $\eta$, as detailed in Appendix \ref{sec:notation}, the AMSE of $\hat{\bm{M}}_o $, which equal $\sum_{ij} \varb_{m,ij}$ is a function of $\eta$. Therefore, we can always select an appropriate $\eta$ to minimize the AMSE. It's noticeable that when we set $\eta = \infty$, the limit distribution of $\hat{m}_{o,ij}$ is equivalent to estimating the matrix $\bm{M}$ independently, which aligns with the theoretical result presented in \cite{IML24} within the regression case of Example \ref{exa:regression}. Therefore, we can theoretically outperform this method.

\subsection{Optimal Selection of $\eta$}
\label{sec:optimal}

% As demonstrated in Theorem \ref{th:qcoOracle} and detailed in Appendix \ref{sec:notation}, the weight parameter $\eta$ plays a pivotal role in the matrices $\bm{\Psi}_j$ and $\tilde{\bm{\Psi}}_j$, significantly influencing the asymptotic variance of the estimator $\hat{m}_{o,ij}$. The careful selection of $\eta$ is crucial for ensuring that the estimator $\hat{\bm{M}}_o$ attains the optimal asymptotic variance. The AMSE can be calculated as follows:

In this section we propose the selection method for $\eta$ to minimize the AMSE of $\hat{\bm{M}}_o $. First, we calculate the AMSE as follows:

\begin{equation}
\label{eq:AMSE}
\begin{aligned}
\text{AMSE} = & \sum_{i,j=1}^{n_1,n_2} \frac{1}{n_2} \begin{pmatrix} \bm{f}_{\star,s,j} \\ \bm{f}_{\star,m,j} \end{pmatrix}^\top \bm{\Phi}_{m,i}^{-1} \tilde{\bm{\Phi}}_{m,i} \bm{\Phi}_{m,i}^{-1} \begin{pmatrix} \bm{f}_{\star,s,j} \\ \bm{f}_{\star,m,j} \end{pmatrix} + \\ 
& \quad \frac{1}{n_1} \sum_{i,j=1}^{n_1,n_2} \begin{pmatrix} \bm{\lambda}_{\star,m1,i} \\ 0 \\ \bm{\lambda}_{\star,m2,i} \end{pmatrix}^\top \bm{\Psi}_j^{-1}(\eta) \tilde{\bm{\Psi}}_j(\eta) \bm{\Psi}_j^{-1}(\eta) \begin{pmatrix} \bm{\lambda}_{\star,m1,i} \\ 0 \\ \bm{\lambda}_{\star,m2,i} \end{pmatrix} \\
= & \sum_{i=1}^{n_1} \tr \left( \bm{\Phi}_{m,i}^{-1} \tilde{\bm{\Phi}}_{m,i} \bm{\Phi}_{m,i}^{-1} \right) + \\
& \quad \frac{1}{n_1} \sum_{j=1}^{n_2} \tr \left( \bm{\Psi}_j^{-1}(\eta) \tilde{\bm{\Psi}}_j(\eta) \bm{\Psi}_j^{-1}(\eta) \left( \bm{\Lambda}_{\star,m1} \ 0 \ \bm{\Lambda}_{\star,m2} \right)^\top \left( \bm{\Lambda}_{\star,m1} \ 0 \ \bm{\Lambda}_{\star,m2} \right) \right),
\end{aligned}
\end{equation}
where $\bm{\Psi}_j(\eta)$ and $\tilde{\bm{\Psi}}_j(\eta)$ represent matrix functions of $\eta$.

The optimal weight parameter \( \eta_{op} \) is identified as the value that minimizes the AMSE. By substituting the matrices with their estimated counterparts and expectations with sample averages in the AMSE formula, and optimizing the sampled AMSE function, we derive the sample optimizer \( \hat{\eta} \). It is noteworthy that under the assumption where $-\mathbb{E}[l^{\prime\prime}(x_{ij}, m_{\star,ij})] = a \mathbb{E}[(l^{\prime}(x_{ij}, m_{\star,ij}))^2]$ for some constant $a$, we have $\eta_{op} = a$. In this case, we can take $\hat{\eta} = \hat{a}$. The derivation of this relationship is provided in Supplement Section 17.
% \ref*{supp-sec:optimal}. 
This scenario is observed in several classical situations, as outlined below:

\begin{enumerate}
\item \textbf{Regression Case:} In the regression scenario presented in Example \ref{exa:regression}, we adopt the loss function $l(x, m) = -\frac{1}{2}(x - m)^2$. Consequently, the expected value of the square of its derivative is given by:
\[
\mathbb{E}[|l^{\prime}(x, m)|^2] = -\sigma^2 l^{\prime\prime}(x, m).
\]
Based on this relationship, we determine the estimator for the weight parameter as $\hat{\eta} = 1/\hat{\sigma}^2$, where $\hat{\sigma}^2$ represents the estimated variance of the error term $\epsilon$:
\[
\hat{\sigma}^2 = \frac{\sum_{i,j=1}^{n_1, n_2} w_{ij} (x_{ij} - \hat{m}_{o,ij})^2}{\sum_{i,j=1}^{n_1, n_2} w_{ij}}.
\]
However, similar to the standard regression case, the estimator is biased because $\mathbb{E}[(x_{ij} - \hat{m}_{o,ij})^2] < \mathbb{E}[(x_{ij} - m_{\star,ij})^2] = \sigma^2$. This bias tends to reduce the variance estimator. To address this issue, we propose a corrected estimator, denoted as $\hat{\sigma}_{co}^2$, which is detailed in Appendix \ref{sec:debiasing}.

\item \textbf{GLM Case:} In the case of a Generalized Linear Model (GLM), as outlined in Example \ref{exa:glm}, we define the response function $l(x, m)$ such that:
\[
l(x_{ij}, m_{ij}) = x_{ij} m_{ij} - b(m_{ij}).
\]
Drawing from the properties of the exponential family, it follows that $b^{\prime\prime}(m) = \frac{1}{\phi} \mathbb{E}[|l^{\prime}(x, m)|^2]$. Consequently, we select the weight parameter $\eta$ to be $ \frac{1}{\hat{\phi}}$, where $\hat{\phi}$ serves as the estimator for the dispersion parameter $\phi$:
\[
\hat{\phi} = \frac{\sum_{i,j=1}^{n_1, n_2} w_{ij} (x_{ij} - b^{\prime}(\hat{m}_{ij}))^2 / b^{\prime\prime}(\hat{m}_{ij})}{\sum_{i,j=1}^{n_1, n_2} w_{ij}}.
\]
\end{enumerate}

\section{Ranks Estimation}
\label{sec:estimationwitoutrank}

When the ranks of the factor matrices \(\bm{F}_{\cdot}\) are unknown a priori, the first step is to estimate these ranks using regularization techniques. Once the ranks are estimated, we can proceed estimating \(\bm{M}\) and \(\bm{\Theta}\), as detailed in Section \ref{sec:estimationwithknownrank} for the case of known ranks.

The determination of the ranks of matrices \(\bm{H}\), \(\bm{M}\), and \(\bm{\Theta}\) can be framed as a rank selection problem. Drawing on parallels from variable selection in regression, we use a nonconvex penalty regularization method to select the true rank of \(\bm{H}_{\star}\):
\begin{equation}
    \label{eq:mcploss}
    \hat{\bm{M}}, \hat{\bm{\Theta}} = \underset{\rank(\bm{H}) \leq k}{\arg\min} \mathcal{L}(\bm{M}, \bm{\Theta}, \eta) + \phi_{\mu}(\bm{H}),
\end{equation}
where \(\mathcal{L}(\bm{M}, \bm{\Theta}, \eta)\) represents the negative pseudo-log-likelihood function \eqref{eq:loglikelihood}, and \(\phi_{\mu}(\bm{H})\) is the matrix nonconvex penalty term. The rank constraint \(k\) is chosen to be a sufficiently large integer, which can increase with the sample size. Specifically, by setting \(k = \min\{2n_1, n_2\}\), this problem is transformed into one without a rank constraint, as discussed in \cite{NNL16, MCN20, SSO24}. Empirically, we choose \(k = \sqrt{\min\{n_1, n_2\}}\), which provides strong statistical guarantees and enhances computational efficiency.

We define \(\phi_{\mu}(\bm{H})\) as the matrix MCP (Minimax Concave Penalty) function:
\begin{equation}
    \label{eq:matmcp}
    \phi_{\mu}(\bm{H}) = \sum_{i = 1}^{\min\{2n_1, n_2\}} \varphi_{\mu}(\sigma_i(\bm{H})),
\end{equation}
where \(\sigma_i(\bm{H})\) denotes the \(i\)-th singular value of \(\bm{H}\), and \(\varphi_{\mu}(\cdot)\) is the MCP function for scalars \cite{MCP10}: $ \varphi_{\mu}(x) = \mu x - \frac{x^2}{2\gamma}$ when $ x \leq \gamma \mu$, and equal $ \frac{\mu^2 \gamma}{2} $ when $ x > \gamma \mu$. 

Here, \(\gamma\) is a regularization parameter greater than 1, and \(\mu\) is a tuning parameter that controls the strength of the penalty. The MCP penalty helps to overcome the rank overestimation issue of nuclear norm regularization, as highlighted in \cite{MCN20}.

The Singular Value Shrinkage Threshold (SVST) algorithm for the MCP estimator \(\hat{\bm{H}} = (\hat{\bm{M}}^{\top}, \hat{\bm{\Theta}})^{\top}\) of \eqref{eq:mcploss} is outlined in Supplement Section 2.
%  \ref*{supp-sec:SVST}.
With \(\hat{\bm{H}}\), we can estimate the ranks \(d_s\), \(d_m\), and \(d_\theta\) using the soft thresholding method:

\begin{algorithm}
\caption{Rank Estimation}
\label{alg:rankselection}
\KwData{The estimator \(\hat{\bm{H}}\)}
\KwResult{The estimators \(\hat{d}_s\), \(\hat{d}_m\), \(\hat{d}_\theta\)}
\KwIn{The threshold \(\mathbf{T} = \mu \gamma\)}
\begin{enumerate}
    \item First, we use the threshold \(\mathbf{T} = \mu \gamma\) from the MCP loss parameter \eqref{eq:matmcp} to obtain \(d = d_s + d_m + d_\theta\), \(d_s + d_m\), and \(d_s + d_\theta\) as follows:
    \begin{equation}
    \label{eq:rankselection}
    \begin{aligned}
    \hat{d} &= \rank(\hat{\bm{H}}), \\
    \widehat{d_s + d_m} &= \operatorname*{card}\{\sigma_i(\hat{\bm{M}}) > \mathbf{T}\}, \\
    \widehat{d_s + d_\theta} &= \operatorname*{card}\{\sigma_i(\hat{\bm{\Theta}}) > \mathbf{T}\},
    \end{aligned}
    \end{equation}
    where \(\operatorname*{card}\{\sigma_i(\hat{\bm{M}}) > \mathbf{T}\}\) is the number of singular values of \(\hat{\bm{M}}\) greater than \(\mathbf{T}\).
    
    \item Estimate the ranks as follows:
    \begin{equation}
    \label{eq:dhat}
    \hat{d}_\theta = \hat{d} - \widehat{d_s + d_m}, \quad
    \hat{d}_m = \hat{d} - \widehat{d_s + d_\theta}, \quad \hat{d}_s = \widehat{d_s + d_m} + \widehat{d_s + d_\theta} - \hat{d}.
    \end{equation}
\end{enumerate}
\end{algorithm}

Regarding the MCP regularization function \(\varphi_{\mu}(\sigma_i(\bm{H}))\) as defined in equation \eqref{eq:matmcp}, the derivative of \(\varphi_{\mu}(\cdot)\) is zero when \(\sigma_i(\bm{H}) \geq \gamma \mu\). This implies that singular values which are large enough are not penalized. Hence, if \(\sigma_i(\hat{\bm{H}}) \geq \mathbf{T}\), this singular value is likely a strong signal from the true matrix \(\bm{H}_\star\), rather than from the estimation error \(\hat{\bm{H}} - \bm{H}_\star\). This leads us to infer that \(\widehat{d_s + d_m}\) and \(\widehat{d_s + d_\theta}\) are effective estimators for their true counterparts. Furthermore, since \(\hat{d}\) is a consistent estimator of \(d\), as we will demonstrate, the estimator in equation \eqref{eq:dhat} is reliable.

After establishing the initial values, we substitute the estimated ranks \(\hat{d}_s\), \(\hat{d}_m\), and \(\hat{d}_\theta\) into the constraint \eqref{eq:shfactormodelmt} within the optimization problem \eqref{eq:Oraclepco} to obtain the two-step estimators \(\hat{\bm{M}}\) and \(\hat{\bm{\Theta}}\). We then proceed with the selection process for \(\eta\), as discussed in Section \ref{sec:optimal}, to refine these estimators and achieve their optimal values.

Moreover, since the MCP estimator \(\hat{\bm{H}}\) provides a satisfactory approximation for \(\bm{H}_\star\), we can use it to determine initial values for \(\bm{\Lambda}_{\cdot}\) and \(\bm{F}_{\cdot}\) in the optimization problem \eqref{eq:Oraclepco}. For further details on the derivation and application of these initial values, please refer to Supplement Section 3.
% \ref*{supp-sec:initialpoint}.

Once the two-step estimators are in place, we proceed to the tuning parameter selection process to determine the value of \(\mu\). In our case, the regularization parameter \(\gamma\) is fixed at 1.5, and the initial weight parameter \(\eta\) is set to 1. The selection of \(\mu\), which plays a critical role in influencing the estimator \(\hat{\bm{H}}\) as defined in equation \eqref{eq:mcploss}, follows the methods outlined in \cite{MCL21} and \cite{QFM21}. Specifically, we adopt an Information Criterion (IC):
\[
\mu_{\text{choose}} = \arg\min_{\mu > 0} \left\{ \mathcal{Q}(\hat{\bm{M}}, \hat{\bm{\Theta}}) + 0.125 \log(n_1 n_2) k_f \right\},
\]
where \(k_f\) represents the degrees of freedom of the model, calculated as:
\[
k_f = \hat{d}_s(2n_1 + n_2 - \hat{d}_s) + (\hat{d}_m + \hat{d}_\theta)(n_1 + n_2 - \hat{d}_m - \hat{d}_\theta),
\]
and the penalty term \(0.125 \log(n_1 n_2)\) adjusts for model complexity, ensuring effective performance in simulations. The term \(\mathcal{Q}(\hat{\bm{M}}, \hat{\bm{\Theta}})\) represents the model's estimation performance and is given by the negative observed log-likelihood. For the normal regression case in Example \ref{exa:regression}, it is:
\[
\mathcal{Q}(\hat{\bm{M}}, \hat{\bm{\Theta}}) = \sum_{i,j} w_{ij} \left\{ -\hat{\theta}_{ij} + \log(\hat{\sigma}) + \frac{\log(2\pi) + 1}{2} \right\} + \log(1 + \exp(\hat{\theta}_{ij})),
\]
where \(\hat{\sigma}\) is the estimated variance of the error term \(\epsilon\), as detailed in Section \ref{sec:optimal}.

For the GLM case illustrated in Example \ref{exa:glm}, with \(\hat{\phi}\) denoting the estimated GLM parameters, \(\mathcal{Q}(\hat{\bm{M}}, \hat{\bm{\Theta}})\) is:
\[
\mathcal{Q}(\hat{\bm{M}}, \hat{\bm{\Theta}}) = \sum_{i,j} w_{ij} \left\{ -\hat{\theta}_{ij} + \frac{x_{ij} \hat{m}_{ij} - b(\hat{m}_{ij})}{\hat{\phi}} + c(x_{ij}, \hat{\phi}) \right\} + \log(1 + \exp(\hat{\theta}_{ij})).
\]

\subsection{Statistical Guarantee}

For convenience of notation, we introduce the Oracle estimator of $\bm{M}, \bm{\Theta}$ as:
\begin{equation}
\label{eq:Oracle} \bm{M}_{o}, \bm{\Theta}_{o} = \underset{\operatorname*{rank}\begin{pmatrix}\bm{M}^{\top}&\bm{\Theta}^{\top}\end{pmatrix} = d}
{\arg\min}\mathcal{L}(\bm{M},\bm{\Theta},\eta),
\end{equation}
where $d$ is the true rank of $(\bm{M}^{\top},\bm{\Theta}^{\top})^{\top}$. Here we also take $\bm{H}_{o}= (\bm{M}_{o}^{\top}, \bm{\Theta}_{o}^{\top})^{\top}$.
\begin{remark}
\label{rm:Oracle} It's noticeable that $\bm{H}_{o}$ is the estimator of known rank estimator \eqref{eq:Oraclepco} when take $d_{s}, d_{m}, d_{\theta}= d, 0, 0$, so that Theorem \ref{th:Oracleproperty} also holds for the Oracle estimator $( \bm{M}_{o}, \bm{\Theta}_{o})$.
\end{remark}

Here we denote $c_{l}= \max\{\eta L_{l}, 1\}$. To show the Oracle property of the MCP estimator $\hat{\bm{H}}$ that $\hat{\bm{H}}= \bm{H}_{o}$, we need the following assumptions:
\begin{assumption*}
\label{as:rankestimation}
\item[{(g)}\label{g}] The true matrix $\bm{H}_{\star}$ is rank $d$, with the estimator $\hat{\bm{H}}$ $\rank(\hat{\bm{H}}) \leq k$, $\| \hat{\bm{M}}\|_{\infty}\leq \alpha_{m}$ and $\|\hat{\bm{\Theta}}\|_{\infty}\leq \alpha_{\theta}$. For $\sigma_{d}(\bm{H}_{\star})$ and $\mu$, under different error term, we need:
\begin{itemize}
\item[{(g1)}\label{g1}]
\[
\sigma_{d}(\bm{H}_{\star}) - \gamma \mu \geq \sqrt{\frac{d \gamma}{\alpha_{h}}}
\mu; \quad
\frac{\min\{\frac{\mu}{2c_l}, \sigma_{d}(\bm{H}_{\star}\} - \gamma
\mu)}{\sqrt{k(n_{1}+ n_{2})}}\frac{\alpha_{h}}{c_{\psi_2}c_{l}}\to
\infty.
\]

\item[{(g2)}\label{g2}]
\[
\sigma_{d}(\bm{H}_{\star}) - \gamma \mu \geq \sqrt{\frac{2 d \gamma}{\alpha_{h}}}
\mu; \quad
\frac{\min\{\frac{\mu}{2c_l}, \sigma_{d}(\bm{H}_{\star}\} - \gamma
\mu)}{\sqrt{k(n_{1}+ n_{2})}}\frac{\alpha_{h}}{ \eta \alpha_{\psi_1}\log(n_{1}n_{2})c_{l}}
\to \infty.
\]

\item[{(g3)}\label{g3}]
\[
\sigma_{d}(\bm{H}_{\star}) - \gamma \mu \geq \sqrt{\frac{2 d \gamma}{\alpha_{h}}}
\mu; \quad \frac{\min\{\frac{\mu}{2c_l}, \sigma_{d}(\bm{H}_{\star}\}
- \gamma \mu)}{k^{1/2}(n_{1}+ n_{2})^{1/2 + 1/(4 + \zeta)}}\frac{\alpha_{h}}{\eta
\alpha_{\zeta}c_{l}}\to \infty.
\]

\item[{(g4)}\label{g4}]
\[
\sigma_{d}(\bm{H}_{\star}) - \gamma \mu \geq \sqrt{\frac{2 d \gamma}{\alpha_{h}}}
\mu; \quad \frac{\min\{\frac{\mu}{2c_l}, \sigma_{d}(\bm{H}_{\star}\}
- \gamma \mu)}{k^{3/4}(n_{1}+ n_{2})^{3/4}}\frac{\alpha_{h}}{\alpha_{m}c_{\delta}c_{l}}
\to \infty.
\]
\end{itemize}
\end{assumption*}

The notations $\alpha_{\cdot}$ are referenced in Assumption {\aref{a} $\sim $ \aref{d}}, and $c_{\cdot}$ are presented in Theorem \ref{th:Oracleproperty}. Assumption \aref{g} imposes constraints on the divergence rate of $\mu$ and $\sigma_{d}(\bm{H}_{\star})$. To satisfy the left side inequality of Assumption \aref{g}, we can select $\mu = o(\sigma_d(\bm{H}_\star))$. Specifically, we present the following lemma to show the lower bound of $\sigma_{d}(\bm{H}_{\star})$:
\begin{lemma}
\label{le:lowerboundsigma}
For shared factor model \eqref{eq:shfactormodelmt}, we have:
\[
\sigma_{d_{s} + d_m + d_\theta}(\bm{H}_{\star}) \geq \sqrt{1 - \xi}\min\{\sigma
_{d_{s} + d_m}(\bm{M}_{\star}), \sigma_{d_{s} + d_\theta}(\bm{\Theta}_{\star})\}.
\]
\end{lemma}

With the common assumption that $\sigma_{d_s + d_m}(\bm{M}_\star), \sigma_{d_s + d_\theta}(\bm{\Theta}_\star) \geq c \sqrt{n_1n_2}$ and Lemma \ref{le:lowerboundsigma}, we can assume $\sigma_d(\bm{H}_\star) \geq c \sqrt{n_1n_2} $. To fulfill the right side inequality of Assumption \aref{g}, for instance, taking \aref{g4} as an example, we can choose $k = O((n_1 + n_2)^{1/6})$ and $\mu = O((n_1 + n_2)^{15/16})$. Consequently, when $n_1 \asymp n_2$, we can comply with Assumption \aref{g4}. Other assumptions can be met using similar approaches. 

We now demonstrate the Oracle property of the MCP estimator $\hat{\bm{H}}$, that $\hat{\bm{H}} = \bm{H}_o $, as presented in the following Theorem:

\begin{theorem}
\label{th:Oracleresult}
For the estimator $\hat{\bm{H}}$ in \eqref{eq:mcploss}, with Assumptions {\aref{a} $\sim$ \aref{d}}, \aref{g}, we have $\hat{\bm{H}}= \bm{H}_{o}$ in probability. The rate of convergence varies depending on the specific assumptions made. Specifically, there exist universal constants $C_{1}, C_{2}, C_{3}, C_{4}$ such that:
\begin{itemize}
\item under Assumptions \aref{d1}, \aref{g1}:
\begin{equation}
\label{eq:Oracleresultd1}
\begin{gathered}
\mathbb{P}(\hat{\bm{H}} \neq \bm{H}_o) \leq C_1 \exp(- C_2 \frac{\alpha_{h}^{2}}{c_{\psi_2}^{2}k}\frac{\mu^{2}}{(n_{1}+
n_{2})c_l^{2}}) + \\ C_1 \exp(- \! C_2 \frac{\alpha_{h}^{2}}{c_{\psi_2}^{2}k}
\frac{(\sigma_{d}(\bm{H}_{\star}) \! - \! \gamma \mu)_{+}^{2}}{n_{1} + n_{2}}) \!
+ \! C_3 \exp(- (C_4 \frac{\mu}{\max\{\eta \alpha_{\psi_2},1\}}
\! - \! \sqrt{n_{1}+ n_{2}})_+^2) \to 0,
\end{gathered}
\end{equation}
where $(a)_{+}$ is the positive part of $a$.

\item under Assumptions \aref{d2}, \aref{g2}:
\begin{equation}
\label{eq:Oracleresultd2}
\begin{gathered}
\mathbb{P}(\hat{\bm{H}} \neq \bm{H}_o) \leq \frac{2}{n_{1}n_{2}} + C_1
\exp(- C_2 \frac{\alpha_{h}^{2}}{\eta^{2}\alpha_{\psi_1}^{2}k}
\frac{\mu^{2}}{(n_{1}+ n_{2})(\log(n_{1}n_{2}))^{2}c_l^{2}}) + \\ C_1 \exp(- C_2
\frac{\alpha_{h}^{2}}{\eta^{2}\alpha_{\psi_1}^{2}k}
\frac{(\sigma_{d}(\bm{H}_{\star}) - \gamma \mu)_{+}^{2}}{(n_{1}+ n_{2})(\log(n_{1}n_{2}))^{2}})
+ \\ C_3 \! \exp(-\! \min\{(C_4 \!
\frac{\mu}{\max\{\eta \alpha_{\psi_1},1\}} \! - \! 
\sqrt{n_{1}+ n_{2}})_{+}^2, \! (C_4 \! \frac{\mu}{\max\{\eta \alpha_{\psi_1},1\}} \!
- \! \sqrt{n_{1}+ n_{2}})_{+}\}) \! \to \! 0.
\end{gathered}
\end{equation}

\item under Assumptions \aref{d3}, \aref{g3}:
\begin{equation}
\label{eq:Oracleresultd3}
\begin{gathered}
\mathbb{P}(\hat{\bm{H}} \neq \bm{H}_o) \leq \frac{n_{1}n_{2}}{(n_{1}+ n_{2})^{2
+ \frac{2\zeta}{16 + \zeta}}} + C_1 \exp(- \!C_2
\frac{\alpha_{h}^{2}}{\eta^{2}\alpha_{\zeta}^{2}k}
\frac{\mu^{2}}{(n_{1}+ n_{2})^{1 + \frac{2}{4 + \zeta / 4}}c_l^{2}}) + \\ C_1 \exp(- \!C_2
\frac{\alpha_{h}^{2}}{\eta^{2}\alpha_{\zeta}^{2}k}
\frac{(\sigma_{d}(\bm{H}_{\star}) - \gamma \mu)_{+}^{2}}{(n_{1}+ n_{2})^{1
+ \frac{2}{4 + \zeta / 4}}}) + C_3 \max\{\eta\alpha_\zeta,1\}\frac{\sqrt{n_{1}+ \!
n_{2}}}{\mu} \to 0.
\end{gathered}
\end{equation}

\item under Assumptions \aref{d4}, \aref{g4}:
\begin{equation}
\label{eq:Oracleresultd4}
\begin{gathered}
\mathbb{P}(\hat{\bm{H}} \neq \bm{H}_o) \leq \frac{n_{1}n_{2}}{(n_{1}+ n_{2})^{2
+ \delta / 2}} + C_1 \exp(- \!C_2
\frac{\alpha_{h}}{c_{\delta}(\alpha_{m}+ \alpha_{\theta}) k^{3/2}}
\frac{\mu^{2}}{(n_{1}+ n_{2})^{3/2}c_l^{2}})
+ \\ C_1 \exp(- \!C_2
\frac{\alpha_{h}}{c_{\delta}(\alpha_{m}+ \alpha_{\theta}) k^{3/2}}
\frac{(\sigma_{d}(\bm{H}_{\star}) - \gamma \mu)_{+}^{2}}{(n_{1}+ n_{2})^{3/2}})
+ C_3 \max\{\eta\alpha_\delta,1\} \frac{\sqrt{n_{1}+ n_{2}}}{\mu} \to 0.
\end{gathered}
\end{equation}
\end{itemize}
\end{theorem}

Note that Assumption \aref{g} merely requires that the rank of $\hat{\bm{H}}$ be less than $k$. This implies that even if we do not explicitly incorporate a rank constraint in the MCP estimator as defined by \eqref{eq:mcploss}, but instead properly select $\mu$ and $\gamma$ to ensure that the estimator $\hat{\bm{H}}$ is low-rank, the theorem remains valid.

According to Theorem \ref{th:Oracleresult}, selecting a smaller value for $k$ diminishes the probability that the estimator $\hat{\bm{H}}$ will deviate from the Oracle estimator $\bm{H}_{o}$. Consequently, with a strong prior on the rank of $\bm{H}_{\star}$, one can choose a smaller $k$ as the constraint of \eqref{eq:mcploss}, which can also accelerate the computation. Additionally, to ensure robust estimation, one can always select $k$ to approach infinity in conjunction with the sample dimensions $n_{1}, n_{2}\to \infty$ as we discussed above.

And for the Oracle estimator $\bm{M}_{o}, \bm{\Theta}_{o}$ and $\bm{H}_{o}$, we take set $\mathcal{A}$ as:
\[
\mathcal{A}= \{ \bm{H}_{o}\text{ is the stationary point of \eqref{eq:mcploss}
and fixed point of function $\mathcal{S}_{k}(\cdot)$ (S1)}\}
% (\ref*{supp-eq:updatek})}\}
\]
where $\mathcal{S}_{k}(\cdot)$ is the one-step update function to optimize \eqref{eq:mcploss}, defined in Supplement Section 2,
% \ref*{supp-sec:SVST}, 
then we have the following lemma:

\begin{lemma}
\label{th:Oraclestationary}
Under Assumptions {\aref{a} $\sim$ \aref{d}}, the probability $\mathbb{P}(\mathcal{A}^{c})$ shares the same probability bound as outlined in Theorem \ref{th:Oracleresult} within equations \eqref{eq:Oracleresultd1} {$\sim $} \eqref{eq:Oracleresultd4} by setting $k = d$.
\end{lemma}
The probability bound of $\mathbb{P}(\mathcal{A}^{c})$ is significantly smaller than it for $\mathbb{P}(\hat{\bm{H}}\neq \bm{H}_{o})$ as $d$ is fixed but $k \to \infty$. Therefore, if we possess prior knowledge that the objective function \eqref{eq:mcploss} has only one stationary point, then the Oracle estimator $(\bm{M}_{o}, \bm{\Theta}_{o})$ becomes the unique solution to \eqref{eq:mcploss}. Under such circumstances, the Assumption \aref{g} is rendered unnecessary.

We observe that when $\hat{\bm{H}}= \bm{H}_{o}$, the total rank estimator $\hat{d}$ from equation \eqref{eq:rankselection} is equivalent to its true value. Consequently, if the estimators $\widehat{d_s + d_m}$ and $\widehat{d_s + d_\theta}$ are accurately determined, then the rank estimators $\hat{d}_{s}$, $\hat{d}_{m}$, and $\hat{d}_{\theta}$ from equation \eqref{eq:dhat} are also correct. This implies that the rank estimation method is consistent. We need additional assumptions for this property's probability bound:
\begin{assumption*}
\item[{(h)}\label{h}] For the true value $\bm{F}_{\star,\cdot}, \bm{\Lambda}_{\star,\cdot}$, we take $\xi$ the fixed bound:
\[
\|\frac{1}{n_{2}}\bm{F}_{\star,m}^{\top}\bm{F}_{\star,\theta}\| \leq \xi < 1.
\]

For true matrix $\bm{M}_{\star}, \bm{\Theta}_{\star}$, the smallest nonzero singular value $\sigma_{d_{s} + d_m}(\bm{M}_{\star}), \sigma_{d_{s} + d_\theta}(\bm{\Theta}_{\star})$ has the following rate:

\begin{itemize}
\item[{(h1)}\label{h1}]
\[
\frac{\min\{\frac{\mu}{2 c_l}, \min\{\sigma_{d_{s} + d_m}(\bm{M}_{\star}),
\sigma_{d_{s} + d_\theta}(\bm{\Theta}_{\star})\} - \gamma \mu\}}{\sqrt{n_{1}+
n_{2}}}\frac{\alpha_{h}}{c_{\psi_2}\sqrt{d}}\to \infty.
\]

\item[{(h2)}\label{h2}]
\[
\frac{\min\{\frac{\mu}{2 c_l}, \min\{\sigma_{d_{s} + d_m}(\bm{M}_{\star}),
\sigma_{d_{s} + d_\theta}(\bm{\Theta}_{\star})\} - \gamma \mu\}}{\sqrt{n_1
+ n_2}(\log(n_{1}) + \log(n_{2}))}\frac{\alpha_{h}}{\eta\alpha_{\psi_1}\sqrt{d}}
\to \infty.
\]

\item[{(h3)}\label{h3}]
\[
\frac{\min\{\frac{\mu}{2 c_l}, \min\{\sigma_{d_{s} + d_m}(\bm{M}_{\star}),
\sigma_{d_{s} + d_\theta}(\bm{\Theta}_{\star})\} - \gamma \mu\}}{(n_{1}+
n_{2})^{1/2 + 1/(4 + \zeta)}}\frac{\alpha_{h}}{\eta \alpha_{\delta}\sqrt{d}}
\to \infty.
\]

\item[{(h4)}\label{h4}]
\[
\frac{\min\{\frac{\mu}{2 c_l}, \min\{\sigma_{d_{s} + d_m}(\bm{M}_{\star}),
\sigma_{d_{s} + d_\theta}(\bm{\Theta}_{\star})\} - \gamma \mu\}}{d^{3/4}(n_{1}+
n_{2})^{3/4}}\frac{\alpha_{h}}{c_{\delta}\alpha_{m}}\to \infty.
\]
\end{itemize}
\end{assumption*}
Similar to Assumption \aref{g}, Assumption \aref{h} governs the divergence rate of the tuning parameter $\mu$ and the smallest non-zero singular value of the matrices $\bm{M}_{\star}$ and $\bm{\Theta}_{\star}$. As we can always consider $\min\{\sigma_{d_s + d_k}(\bm{M}_\star), \sigma_{d_s + d_\theta}(\bm{\Theta}_\star)\} $ is the order $\sqrt{n_1n_2} $, we can select \(\mu\) accordingly to satisfy Assumption \aref{h}. For instance, drawing from the discussion on Assumption \aref{g}, we can opt for \(\mu = O((n_1 + n_2)^{15/16})\).

We define the number \(\mathcal{B}\) as:
\[
\mathcal{B} = \min\left\{ \gamma \mu, \min\{\sigma_{d_{s} + d_m}(\bm{M}_{\star}), \sigma_{d_{s} + d_\theta}(\bm{\Theta}_{\star})\} - \gamma \mu\right\} \to \infty.
\]

Then for the ranks estimation, we have the following nonasymptotic bound:
% As demonstrated in the Supplementary material's Section \ref*{supp-subsec:proofdresult}, when \(\hat{\bm{H}} = \bm{H}_{o}\) and \(\|\bm{H}_{o} - \bm{H}_{\star}\|_{F} \leq \mathcal{B}\), the estimators \(\hat{d}_{s}, \hat{d}_{m}, \hat{d}_{\theta}\) are accurate. Consequently, we present the following theorem:
\begin{theorem}
\label{th:dresult}

For estimators of $d, d_{m}, d_{\theta}$ from \eqref{eq:dhat}, with Assumption $\aref{a} {\sim} \aref{d}, \aref{g}, \aref{h}$, we have the probability bound:
\[
\mathbb{P}(\hat{d}_{s}, \hat{d}_{m}, \hat{d}_{\theta}\neq d_{s}, d_{m}, d
_{\theta}) \leq \mathbb{P}(\hat{\bm{H}}\neq \bm{H}_{o}) + \mathbb{P}( \|\bm{H}_{o}- \bm{H}_{\star}
\|_{F}\geq \mathcal{B}) \to 0.
\]

The nonasymptotic form on the right-hand side can be derived from Theorems \ref{th:Oracleproperty}, \ref{th:Oracleresult} and the discussion on $\bm{H}_{o}$ in Remark \ref{rm:Oracle}. Under our assumptions, this term will tend towards zero.
\end{theorem}
 
Theorem \ref{th:dresult} shows that the rank estimators \( \hat{d}_{s}, \hat{d}_{m}, \hat{d}_{\theta} \) are consistent, which is essential for determining the dimensions of both the shared and specific factors for shared factor model \eqref{eq:shfactormodel}.

\section{Simulation}
\label{sec:simulation}

In this section, we assess the empirical performance of the proposed shared factor based estimator as detailed in Sections \ref{sec:estimationwitoutrank} and \ref{sec:estimationwithknownrank}. Our experiments are designed to explore the accuracy of the estimator under settings of shared and specific factors, error term distributions, and observation rates.

Given sample size $n = n_{1}= n_{2}$ and ranks $d_{s}, d_{m}, d_{\theta}$, we generate the factor and factor loading matrices as follows:
\begin{equation*}
    \label{eq:simulationdata}
    \begin{gathered}
        \bm{\Lambda}_{m2} = [\mathcal{N}(0, \frac{1}{d_s + d_m - 1} )]_{n\times d_m}, \quad
        \bm{\Lambda}_{m1} =
    \begin{pmatrix}
    \bm{1}_{n} & [\mathcal{N}(0, \frac{1}{d_s + d_m - 1})]_{n \times (d_s - 1)}
    \end{pmatrix}, \\ 
    \bm{\Lambda}_{\theta_2} = [\mathcal{N}(0,\frac{1}{d_s + d_\theta - 1} )]_{n\times d_m}, \quad \bm{\Lambda}_{\theta 1} =
        \begin{pmatrix}
        - m_{p}\bm{1}_{n} & [\mathcal{N}(0, \frac{1}{d_s + d_\theta - 1})]_{n \times (d_s - 1)}
        \end{pmatrix}, \\ 
        \bm{F}_m = [\mathcal{N}(0,1)]_{n\times d_m}, \quad \bm{F}_\theta = [\mathcal{N}(0,1)]_{n\times d_\theta}, \quad \bm{F}_s =
            \begin{pmatrix}
            \bm{1}_{n} & [\mathcal{N}(0, 1)]_{n \times (d_s - 1)}
            \end{pmatrix},
    \end{gathered}
\end{equation*}
where, \( [\mathcal{N}(0, \sigma^2)]_{m \times n} \) denotes an \( m \times n \) random matrix with each element independently sampled from the mean zero normal distribution with variance $\sigma^2 $, and \( \bm{1}_{n} \) represents the \( n \times 1 \) vector with all elements equal to \( 1 \). According to the model in \eqref{eq:shfactormodelmt}, the matrices \( \bm{M} \) and \( \bm{\Theta} \) are calculated as:
\[
\bm{M} = \bm{\Lambda}_{m1} \bm{F}_s^{\top} + \bm{\Lambda}_{m2} \bm{F}_m^{\top}, \quad \bm{\Theta} = \bm{\Lambda}_{\theta 1} \bm{F}_s^{\top} + \bm{\Lambda}_{\theta 2} \bm{F}_\theta^{\top}.
\]

This configuration ensures that the variance of each element in matrices $\bm{M}$ and $\bm{\Theta}$ is 1, with means of 1 and $m _{p}$, respectively. And $m_{p}$ is a parameter utilized to control the missing probability: setting $m_{p}= 1$ results in an observation rate of approximately 30\%. By increasing $m_{p}$ to 1.5 and 2, the observation rates are reduced to about 22\% and 15\%, respectively.

In this context, we focus solely on the regression model presented in Example \ref{exa:regression}. That with parameters $\bm{M}, \bm{\Theta}$, we generate the data matrix $\bm{X}$ and the missing matrix $\bm{W}$ as follows:
\[
w_{ij}\sim \mathcal{B}(\expit(\theta_{ij})), \quad x_{ij}= m_{ij}+ \epsilon_{ij}
,
\]
where $w_{ij}$ is drawn from a Bernoulli distribution with probability $\expit(\theta_{ij})$, and $x_{ij}$ represents the observed data with error term $\epsilon_{ij}$. The error term $\epsilon_{ij}$ is selected from mean zero normal distributions with variances of 0.5, 1, and 1.5.
%  and from a standardized Student's t-distribution with degrees of freedom 5 and 9. This selection provides a robust framework for assessing the model's performance under various error assumptions.

We set the sample size $n = n_{1}= n_{2}$ to both 500 and 1000, fix the total rank of the matrix $\bm{H}$ at 9, and explore three distinct cases for the rank configuration $(d_{s}, d_{\theta}, d_{m})$: \textbf{(i) Balanced Case Scenario:} We analyze the impact of varying the number of shared factors on estimator accuracy. We assign the tuples $(d_{s}, d_{\theta}, d_{m})$ the following values: $(9, 0, 0)$, $(5, 2, 2)$, and $(1, 4, 4)$, each representing a different scenario with an equal number of specific factors for matrices $\bm{M}$ and $\bm{\Theta}$. \textbf{(ii) Specific Case for $\bm{M}$:} We examine scenarios where only matrix $\bm{M}$ contains specific factors, setting $(d_{s}, d_{\theta}, d_{m})$ to $(7, 0, 2)$, $(5, 0, 4)$, and $(3, 0, 6)$. The goal is to understand how the proportion of shared factors affects the accuracy of $\bm{M}$'s estimation. \textbf{(iii) Specific Case for $\bm{\Theta}$:} Similarly, we set $(d_{s}, d_{\theta}, d_{m})$ to $(7, 2, 0)$, $(5, 4, 0)$, and $(3, 6, 0)$ to study the influence of shared factors on the accuracy of $\bm{\Theta}$'s estimation. % \subsection*{Performance Evaluation Metrics}

We denote our shared factor-based two-step estimator with optimal selection of $\eta$ as OSH (Optimal Shared Factor), and the estimator with $\eta = 1$ as SH (Shared Factor). For comparative purposes, we evaluate several existing methods under the MAR model with IPW estimators, including MHT \cite{MHT10}, NW \cite{NW12} and MWC \cite{MCL21}, each with its missing mechanism assumption:

\begin{enumerate}
\item MWC: The missing probability for an element is determined by a low-rank
matrix.

\item MHT: The missing probability is constant, indicating an missing complete at random (MCAR) mechanism.

\item NW: The missing probability for an element is the product of the
corresponding row and column missing probabilities.
\end{enumerate}

The details regarding the estimator of the missing probability and the target matrix for comparative methods are provided in Supplement Section 20. 
% \ref*{supp-sec:simulationresults}. 
It is important to note that the methods presented in the referenced articles utilize matrix nuclear norm regularization to enforce a low-rank structure, which may result in biased matrix estimation. To ensure a fair comparison in our simulation results, we assume that the rank of the low-rank matrices is known for the comparative methods, using the Oracle estimator for comparison instead of the nuclear penalized estimator.

In this section, we focus on the two performance metrics: \textbf{(i)Estimation Accuracy:} We evaluate the accuracy of $\hat{\bm{M}}$ by the mean squared error (MSE): $\mse(\hat{\bm{M}}) = \frac{1}{n_{1}n_{2}}\|\hat{\bm{M}}- \bm{M}_{\star}\|_{F}^{2}$. And to compare the OSH estimator with the best comparative method, we calculate the ratio: $\mathrm{Ratio}_{1}= 1 - \frac{\mse(\hat{\bm{M}}_{OSH})}{\mse(\hat{\bm{M}}_{MHT})}.$ \textbf{(ii)Rank Estimation Accuracy:} We assess the accuracy of the estimated ranks using the proportion of the estimated rank to the true rank. For the comparison of $\hat{\bm{\Theta}}$'s MSE, the improvement of optimal weight selection procedure, the sampled MSE vs theoretical AMSE, and the behavior of $\hat{\bm{M}}$ under more heavy-tailed error terms, we refer to the Supplement Section 20.
% \ref*{supp-sec:simulationresults}.

\subsection{Compare with Existing Methods}

Due to space constraints, we focus our comparison on the estimation of $\hat{\bm{M}}$ with the MHT method, as it's the best performing comparative method. We only present a subset of the $(d_{s}, d_{\theta}, d_{m})$ configurations here, with the complete results provided in Supplement Section 20.1.
%  \ref*{supp-sec:compare}.

\begin{table}
\caption{Simulation $\mse(\hat{\bm{M}})$ results when
$\epsilon \sim \mathcal{N}(0,\sigma^{2})$, sample size $n = 500$. The first
row of data is the mean value, and the second row is the standard deviation.}
\label{tab:simulationofM}
\begin{adjustbox}{width = 400pt}
\begin{tabular}{ccccccccccc}
\hline
\tiny{$\epsilon =\mathcal{N}(0,\sigma^{2})$ } & Model  & OSH    & SH& MHT   & OSH      & SH       & MHT      & OSH      & SH       & MHT      \\
$\sigma^{2}$ & $d_{s}, d_{\theta}, d_{m}$ & \multicolumn{3}{c}{30\% observation rate} & \multicolumn{3}{c}{22\% observation rate} & \multicolumn{3}{c}{15\% observation rate} \\
\hline
\multirow{6}{*}{0.5} & \multirow{2}{*}{5,2,2} & \textbf{0.0484} & 0.0506 & 0.0531 & \textbf{0.0694} & 0.0727 & 0.0786 & \textbf{0.1067} & 0.1117 & 0.1281 \\
 &  & (0.0008) & (0.0009) & (0.0009) & (0.0014) & (0.0015) & (0.0017) & (0.0028) & (0.0031) & (0.0030) \\
 & \multirow{2}{*}{5,0,4} & \textbf{0.0634} & 0.0654 & 0.0715 & \textbf{0.0929} & 0.0959 & 0.1086 & \textbf{0.1478} & 0.1522 & 0.1980 \\
 &  & (0.0012) & (0.0013) & (0.0012) & (0.0020) & (0.0020) & (0.0025) & (0.0038) & (0.0040) & (0.0358) \\
 & \multirow{2}{*}{5,4,0} & \textbf{0.0341} & 0.0361 & 0.0368 & \textbf{0.0478} & 0.0509 & 0.0532 & \textbf{0.0717} & 0.0765 & 0.0834 \\
 &  & (0.0007) & (0.0008) & (0.0008) & (0.0012) & (0.0013) & (0.0014) & (0.0020) & (0.0020) & (0.0022) \\
 \\
\multirow{6}{*}{1} & \multirow{2}{*}{5,2,2} & \textbf{0.0937} & 0.0937 & 0.1091 & \textbf{0.1347} & 0.1348 & 0.1637 & \textbf{0.2091} & 0.2092 & 0.2720 \\
 &  & (0.0016) & (0.0016) & (0.0018) & (0.0029) & (0.0029) & (0.0032) & (0.0059) & (0.0059) & (0.0076) \\
 & \multirow{2}{*}{5,0,4} & \textbf{0.1246} & 0.1246 & 0.1492 & \textbf{0.1849} & 0.1849 & 0.2345 & \textbf{0.3001} & 0.3001 & 0.4265 \\
 &  & (0.0026) & (0.0026) & (0.0027) & (0.0040) & (0.0040) & (0.0214) & (0.0084) & (0.0084) & (0.0272) \\
 & \multirow{2}{*}{5,4,0} & \textbf{0.0653} & 0.0653 & 0.0747 & \textbf{0.0912} & 0.0912 & 0.1088 & \textbf{0.1371} & 0.1371 & 0.1709 \\
 &  & (0.0014) & (0.0014) & (0.0016) & (0.0023) & (0.0023) & (0.0028) & (0.0040) & (0.0040) & (0.0045) \\
 \\
\multirow{6}{*}{1.5} & \multirow{2}{*}{5,2,2} & \textbf{0.1384} & 0.1394 & 0.1681 & \textbf{0.2006} & 0.2020 & 0.2555 & \textbf{0.3175} & 0.3194 & 0.4344 \\
 &  & (0.0025) & (0.0025) & (0.0027) & (0.0045) & (0.0045) & (0.0052) & (0.0093) & (0.0093) & (0.0130) \\
 & \multirow{2}{*}{5,0,4} & \textbf{0.1873} & 0.1883 & 0.2328 & \textbf{0.2822} & 0.2837 & 0.3694 & \textbf{0.4749} & 0.4767 & 0.7037 \\
 &  & (0.0040) & (0.0040) & (0.0043) & (0.0061) & (0.0061) & (0.0076) & (0.0155) & (0.0159) & (0.0239) \\
 & \multirow{2}{*}{5,4,0} & \textbf{0.0950} & 0.0958 & 0.1137 & \textbf{0.1329} & 0.1339 & 0.1667 & \textbf{0.2093} & 0.2105 & 0.2651 \\
 &  & (0.0021) & (0.0021) & (0.0025) & (0.0035) & (0.0035) & (0.0042) & (0.0245) & (0.0243) & (0.0068) \\
\hline
\end{tabular}
\end{adjustbox}
\end{table}

\begin{table}
\centering
\caption{Sample $\operatorname{Ratio}_1$ value for different missing rates and models when  $n = 500 $, the first row of data is the mean value, and the second row is the standard deviation.}
\label{tab:ratio1}
\begin{adjustbox}{width = 400pt}
\begin{tabular}{cccccccccc}
\hline 
Model & \multicolumn{3}{c}{$m_p = 1 $ observed rate $30\% $} & \multicolumn{3}{c}{$m_p = 1.5 $ observed rate $22\% $} & \multicolumn{3}{c}{$m_p = 2 $ observed rate $15\% $} \\
$d_{s},\! d_{\theta},\! d_{m}$ & \tiny{$\var(\epsilon) = 0.5 $} &   \tiny{$\var(\epsilon) = 1 $}  & \tiny{$\var(\epsilon) = 1.5$} & \tiny{$\var(\epsilon) = 0.5 $} &   \tiny{$\var(\epsilon) = 1 $}  & \tiny{$\var(\epsilon) = 1.5$} & \tiny{$\var(\epsilon) = 0.5 $} &   \tiny{$\var(\epsilon) = 1 $}  & \tiny{$\var(\epsilon) = 1.5$} \\
\hline
\multirow{2}{*}{9,0,0} & 0.1288 & 0.2121 & 0.2644 & 0.1769 & 0.2670 & 0.3224 & 0.2467 & 0.3707 & 0.4345 \\
& (0.0107) & (0.0094) & (0.0089) & (0.0643) & (0.0106) & (0.0092) & (0.0518) & (0.0406) & (0.0342) \\
\multirow{2}{*}{5,2,2} & 0.0879 & 0.1409 & 0.1768 & 0.1162 & 0.1770 & 0.2149 & 0.1665 & 0.2309 & 0.2690 \\
& (0.0098) & (0.0088) & (0.0088) & (0.0155) & (0.0136) & (0.0131) & (0.0160) & (0.0175) & (0.0154) \\
\multirow{2}{*}{1,4,4} & 0.0263 & 0.0362 & 0.0421 & 0.0492 & 0.0592 & 0.0627 & 0.0813 & 0.0781 & 0.0756 \\
& (0.0075) & (0.0069) & (0.0061) & (0.0108) & (0.0083) & (0.0076) & (0.0118) & (0.0118) & (0.0090) \\
\multirow{2}{*}{7,0,2} & 0.1215 & 0.1898 & 0.2340 & 0.1603 & 0.2392 & 0.2867 & 0.2328 & 0.3387 & 0.3832 \\
& (0.0088) & (0.0090) & (0.0083) & (0.0131) & (0.0130) & (0.0120) & (0.0183) & (0.0498) & (0.0169) \\
\multirow{2}{*}{5,0,4} & 0.1125 & 0.1648 & 0.1957 & 0.1444 & 0.2077 & 0.2359 & 0.2393 & 0.2943 & 0.3248 \\
& (0.0094) & (0.0091) & (0.0087) & (0.0140) & (0.0448) & (0.0115) & (0.0820) & (0.0367) & (0.0201) \\
\multirow{2}{*}{3,0,6} & 0.0894 & 0.1156 & 0.1316 & 0.1268 & 0.1536 & 0.1651 & 0.1947 & 0.2473 & 0.2516 \\
& (0.0090) & (0.0071) & (0.0072) & (0.0703) & (0.0488) & (0.0126) & (0.0551) & (0.0807) & (0.0647) \\
\multirow{2}{*}{7,2,0} & 0.0993 & 0.1671 & 0.2141 & 0.1322 & 0.2109 & 0.2556 & 0.1917 & 0.2692 & 0.3191 \\
& (0.0121) & (0.0118) & (0.0113) & (0.0141) & (0.0146) & (0.0357) & (0.0665) & (0.0193) & (0.0169) \\
\multirow{2}{*}{5,4,0} & 0.0734 & 0.1259 & 0.1644 & 0.1012 & 0.1619 & 0.2025 & 0.1402 & 0.1976 & 0.2104 \\
& (0.0107) & (0.0104) & (0.0100) & (0.0160) & (0.0143) & (0.0131) & (0.0173) & (0.0159) & (0.0930) \\
\multirow{2}{*}{3,6,0} & 0.0586 & 0.0976 & 0.1257 & 0.0914 & 0.1367 & 0.1660 & 0.1335 & 0.1692 & 0.1861 \\
& (0.0104) & (0.0103) & (0.0097) & (0.0135) & (0.0120) & (0.0116) & (0.0175) & (0.0159) & (0.0495) \\
\hline
\end{tabular}
\end{adjustbox}
\end{table}

Tables \ref{tab:simulationofM} and \ref{tab:ratio1} illustrate the superior performance of the OSH estimator, and demonstrate a lower MSE for $\bm{M}$'s estimator compared to competing methods. The results show a clear trend: as the observation rate decreases, the improvement of the OSH method is more pronounced. For instance, take $\epsilon \sim \mathcal{N}(0,1)$ as an example, when 30\% observation rate, the OSH method shows an improvement of 4\% to 21\% over the MHT method across various models. With 22\% of the data observed, the improvement in $\hat{\bm{M}}$'s MSE ranges from 6\% to 27\%. This enhancement becomes even more pronounced when $m_{p}$ is set to 2, resulting in only 15\% of the data observed, with the improvement range extending to 8\% to 37\%. This finding underscores the importance of incorporating the shared factor structure into our estimator, especially in scenarios with sparser data.

Also, as the variance of $\epsilon$ increases, the OSH estimator exhibits a more favorable performance compared to other methods. For instance, when the observation rate is 30 \% and the variance of $\epsilon$ is 0.5, the improvement in $\hat{\bm{M}}$'s MSE ranges from 3\% to 13\%. This range broadens to 4\% to 26\% when the variance of $\epsilon$ is 1.5. These results indicate that for larger error terms, the OSH estimator is more robust than other methods.
    
Additionally, as shown in Table \ref{tab:simulationofM}, the MSE for $\bm{M}$ is reduced by employing the optimization of the $\eta$ selection procedure when $\var(\epsilon) \neq 1$, and yield identical results when $\var(\epsilon) = 1$. This reduction is particularly pronounced when the shared factor component plays a more significant role, as indicated by an increase in the ratio $d_{s}/ (d_{s}+ d_{m})$. 

\subsection{Rank Estimation Accuracy}
    
To evaluate the accuracy of rank estimation by the OSH estimator, we take our tuning parameter selection procedural and conduct a comprehensive analysis. This involved calculating the ratio of the estimated rank to the true rank over 50 simulations, and the outcomes of this thorough investigation are compiled in Table \ref{tab:ic}.

\begin{table}
\centering
\caption{Rank selection right rate for different missing rates and models}
\label{tab:ic}
\begin{adjustbox}{width = 400pt}
\begin{tabular}{cccccccccc}
\hline
Model & \multicolumn{3}{c}{$m_{p}= 1$ observed rate $30\%$} & \multicolumn{3}{c}{$m_{p}= 1.5$ observed rate $22\%$} & \multicolumn{3}{c}{$m_{p}= 2$ observed rate $15\%$} \\
$d_{s},\! d_{\theta},\! d_{m}$ & \tiny{$\var(\epsilon) = 0.5$}   & \tiny{$\var(\epsilon) = 1$}       & \tiny{$\var(\epsilon) = 1.5$}  & \tiny{$\var(\epsilon) = 0.5$} & \tiny{$\var(\epsilon) = 1$} & \tiny{$\var(\epsilon) = 1.5$} & \tiny{$\var(\epsilon) = 0.5$} & \tiny{$\var(\epsilon) = 1$} & \tiny{$\var(\epsilon) = 1.5$} \\
\hline
\multicolumn{10}{c}{Accurately estimated rate for $n_{1}= n_{2}= 500$}  \\
9,0,0 & 100.0\%& 100.0\%  & 100.0\%    & 100.0\%   & 100.0\% & 100.0\%   & 100.0\%   & 100.0\% & 100.0\%   \\
5,2,2 & 100.0\%& 100.0\%  & 100.0\%    & 100.0\%   & 100.0\% & 100.0\%   & 100.0\%   & 100.0\% & 100.0\%   \\
1,4,4 & 100.0\%& 100.0\%  & 100.0\%    & 100.0\%   & 100.0\% & 100.0\%   & 100.0\%   & 100.0\% & 100.0\%   \\
7,0,2 & 100.0\%& 100.0\%  & 100.0\%    & 100.0\%   & 100.0\% & 100.0\%   & 100.0\%   & 100.0\% & 100.0\%   \\
5,0,4 & 100.0\%& 100.0\%  & 100.0\%    & 100.0\%   & 100.0\% & 100.0\%   & 100.0\%   & 100.0\% & 100.0\%   \\
3,0,6 & 100.0\%& 100.0\%  & 100.0\%    & 100.0\%   & 100.0\% & 100.0\%   & 100.0\%   & 98.0\%  & 94.0\%    \\
7,2,0 & 100.0\%& 100.0\%  & 100.0\%    & 100.0\%   & 100.0\% & 98.0\%    & 100.0\%   & 100.0\% & 98.0\%    \\
5,4,0 & 100.0\%& 100.0\%  & 100.0\%    & 100.0\%   & 100.0\% & 100.0\%   & 100.0\%   & 94.0\%  & 86.0\%    \\
3,6,0 & 100.0\%& 100.0\%  & 100.0\%    & 100.0\%   & 100.0\% & 100.0\%   & 98.0\%    & 92.0\%  & 76.0\%    \\
\multicolumn{10}{c}{Accurately estimated rate for $n_{1}= n_{2}= 1000$} \\
9,0,0 & 100.0\%& 100.0\%  & 100.0\%    & 100.0\%   & 100.0\% & 100.0\%   & 100.0\%   & 100.0\% & 100.0\%   \\
5,2,2 & 100.0\%& 100.0\%  & 100.0\%    & 100.0\%   & 100.0\% & 100.0\%   & 100.0\%   & 100.0\% & 100.0\%   \\
1,4,4 & 100.0\%& 100.0\%  & 100.0\%    & 100.0\%   & 100.0\% & 100.0\%   & 100.0\%   & 100.0\% & 100.0\%   \\
7,0,2 & 100.0\%& 100.0\%  & 100.0\%    & 100.0\%   & 100.0\% & 100.0\%   & 100.0\%   & 100.0\% & 100.0\%   \\
5,0,4 & 100.0\%& 100.0\%  & 100.0\%    & 100.0\%   & 100.0\% & 100.0\%   & 100.0\%   & 100.0\% & 100.0\%   \\
3,0,6 & 100.0\%& 100.0\%  & 100.0\%    & 100.0\%   & 100.0\% & 100.0\%   & 100.0\%   & 100.0\% & 100.0\%   \\
7,2,0 & 100.0\%& 100.0\%  & 100.0\%    & 100.0\%   & 100.0\% & 100.0\%   & 100.0\%   & 100.0\% & 100.0\%   \\
5,4,0 & 100.0\%& 100.0\%  & 100.0\%    & 100.0\%   & 100.0\% & 100.0\%   & 100.0\%   & 100.0\% & 100.0\%   \\
3,6,0 & 100.0\%& 100.0\%  & 100.0\%    & 100.0\%   & 100.0\% & 100.0\%   & 100.0\%   & 100.0\% & 100.0\%   \\
\hline
\end{tabular}
\end{adjustbox}
\end{table}

The results depicted in Table \ref{tab:ic} reveal a clear trend: the precision of rank estimation is decreased with an increase in the specific matrix's rank proportion, the error term's variance, or the missing data rate. 

For example, with a sample size of 500, the poorest rank selection accuracy is encountered in the model with $d_{s}= 3, d_{\theta}= 6, d_{m}= 0$, exhibiting an accuracy rate of 76\% for $\var(\epsilon) = 1.5$ and an observation rate of 15\%. Nevertheless, the findings also suggest that these challenges can be significantly addressed by enlarging the sample size. Specifically, doubling the row and column significantly enhances the rank selection accuracy, corroborating the theoretical predictions.

\section{Real Data Analysis}
\label{sec:realdata}

The development of recommendation systems is a pivotal application within the domain of matrix completion, which is based on estimating users' preferences and items' characteristics to fill in the missing values and offer personalized suggestions to users. By detailed modeling of the missing data mechanism, we can improve the accuracy of the imputed values, which enables the provision of more personalized and precise recommendations to users, thereby enhancing the overall effectiveness of the recommendation system.

% In this section, we conduct an empirical analysis using the MovieLens 100k database \cite{mld16}, a dataset widely recognized in the field of recommender systems. This dataset includes 943 users, 1682 movies, and 100,000 ratings collected over seven months, from September 19th, 1997, to April 22nd, 1998. The ratings are scaled from 1 to 5, with an average rating of 3.53 and a variance of 1.27, reflecting a broad spectrum of user preferences.  The observation rate for each user varies from 0.012 to 0.438, and for each movie, it ranges from 0.001 to 0.618, with an overall observation rate of 0.063. We present the histogram of ratings and cumulative distribution function (CDF) of the rated number of movies for both user margin and movie margin in Figure \ref{fig:user_movie}. 

\begin{figure}
\centering
\subfigure[Histogram of ratings.]{ \includegraphics[width=120pt]{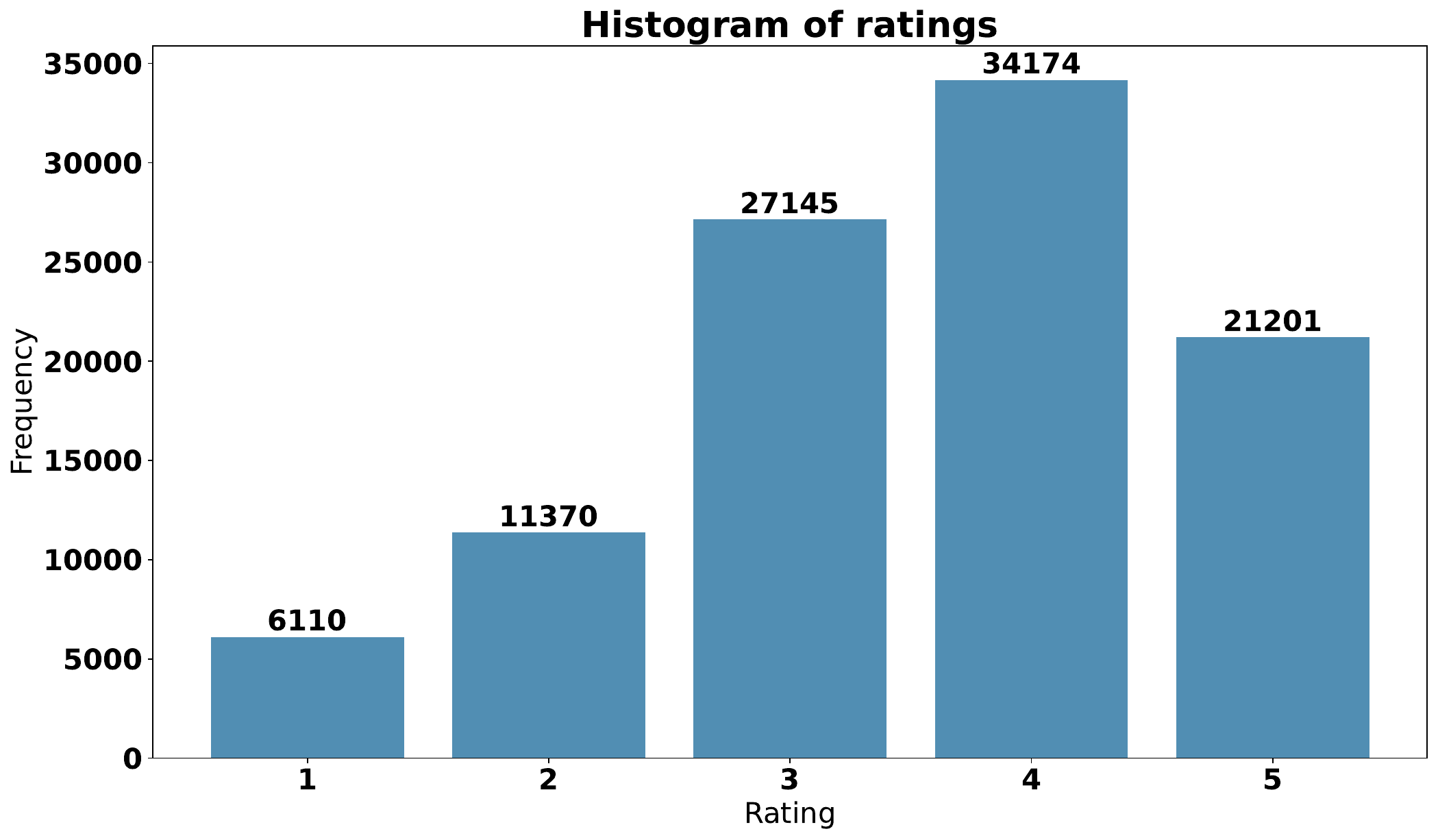} }
\subfigure[CDF of number of rating per user.]{ \includegraphics[width=120pt]{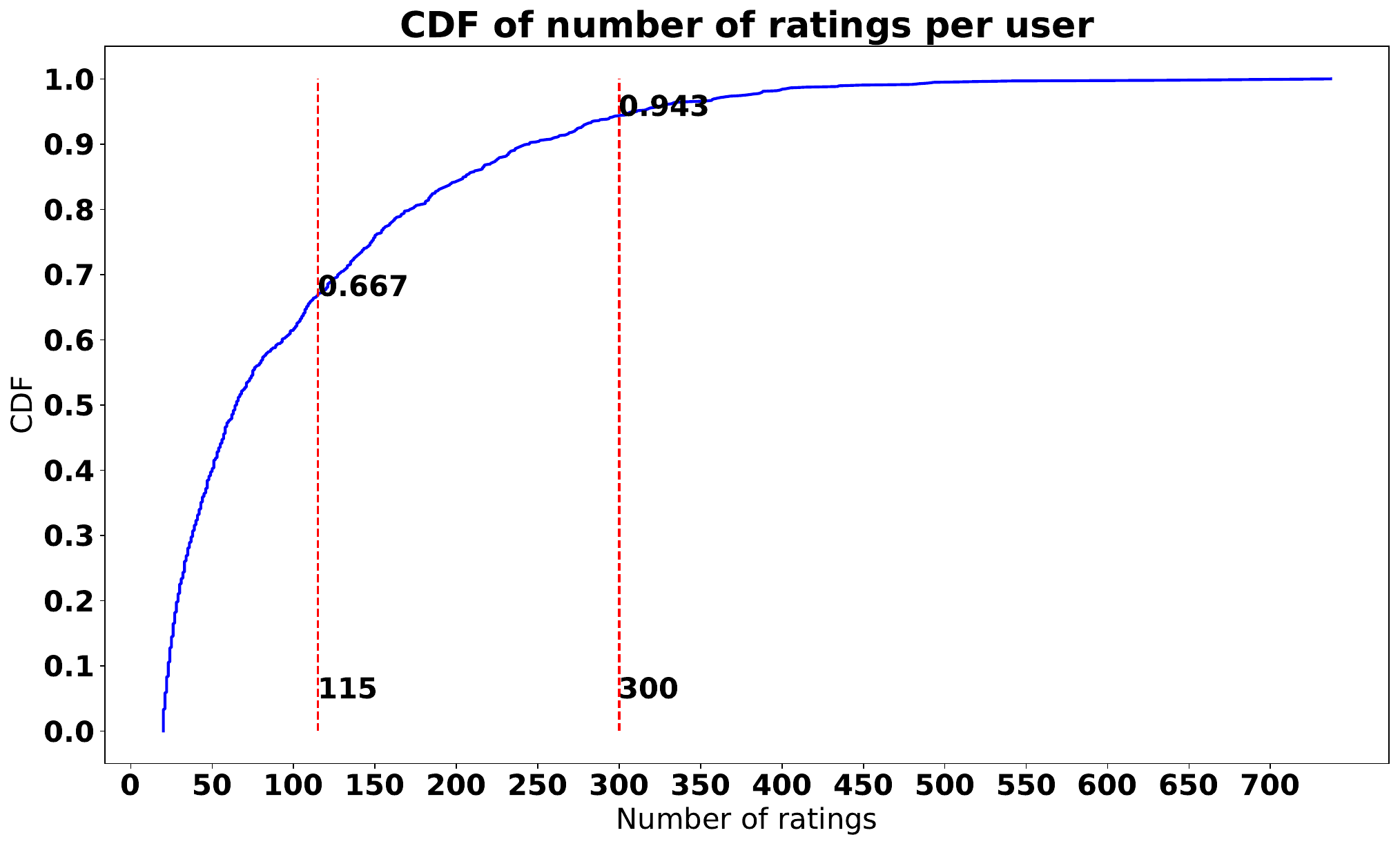} }
\subfigure[CDF of number of rating per movie.]{ \includegraphics[width=120pt]{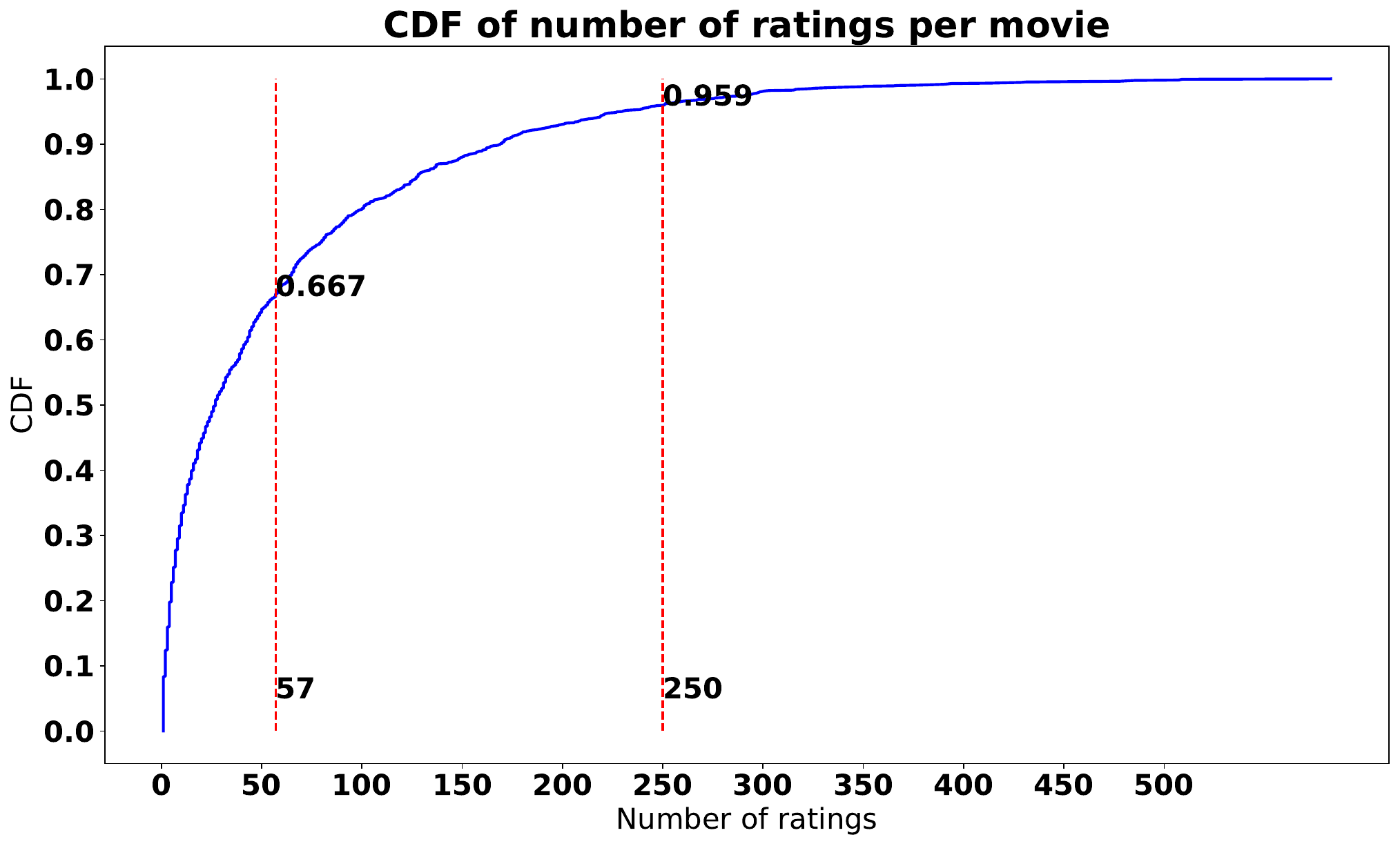} }
\caption{The Ratings Histogram and Cumulate distribution function (CDF) of
user's and movie's rating number correspondingly.}
\label{fig:user_movie}
\end{figure}

In this section, we conduct an empirical analysis using the MovieLens 100k database \cite{mld16}, a dataset widely recognized in the field of recommender systems. This dataset includes 943 users, 1682 movies, and 100,000 ratings collected over seven months, from September 19th, 1997, to April 22nd, 1998. The ratings are scaled from 1 to 5, with an average rating of 3.53 and a variance of 1.27, reflecting a broad spectrum of user preferences. The observation rate for each user varies from $0.012$ to $0.438$, nearly one-third of the users have rated more than 115 movies, with most users have rated fewer than 300 movies. For each movie, the rate ranges from $0.001$ to $0.618$, nearly one-third have been rated by more than 57 users, and most movies have been rated by fewer than 250 users. The overall observation rate is $0.063$. We present the histogram of ratings and cumulative distribution function (CDF) of the rated number of movies for both user margin and movie margin in Figure \ref{fig:user_movie}. 

Before employing our model for analysis, we first assess the relationship between the target and the missing probability parameter matrix's row space--the linear space spanned by their factors correspondingly. We separately estimate $\bm{M}$ and $\bm{\Theta}$, using the MCP term to regularize the matrix estimation and show the result in Supplement Section 21. 
% \ref*{supp-sec:arealdata}. 
Through our analysis, we determine that the ranks of $\hat{\bm{M}}$ and $\hat{\bm{\Theta}}$ are $6$ and $7$, respectively.

Here we introduce a distance measure, denoted as $d_{k}(\cdot,\cdot)$, to quantify the separation between the most significant $k$-dimensional subspaces of the two vector spaces: For $n \times d_{1}$ and $n \times d_{2}$ orthogonal matrices $\bm{V}_{1}$ and $\bm{V}_{2}$, where $d_{1} \leq d_{2}$, and given the singular values $\{\sigma_{i} \}_{i=1}^{d_1}$ of $\bm{V}_{1}^{\top} \bm{V}_{2}$ listed in descending order, the $d_{k}(\bm{V}_1,\bm{V}_2)$ is defined as:
\[
d_{k}(\bm{V}_{1},\bm{V}_{2}) = \sqrt{ \frac{1}{k}\sum_{i=1}^{k} (1 - \sigma_{i}^{2}) },
\]
which represents a kind of average of the sin values for vectors originating from the two spaces. And the correlation between the $6$-dimensional subspaces is evidenced by the measure $\sqrt{1 - d_{k}^{2}(\row(\hat{\bm{M}}),\row(\hat{\bm{\Theta}}))}$, which yields a value of 0.788. This value indicates a substantial degree of correlation, implying a shared factor structural between $\bm{M}$ and $\bm{\Theta}$.

Now, we present the prediction performance of our method. We utilize the pre-split data files ranging from 'u1.base' and 'u1.test' to 'u5.base' and 'u5.test' available on the MovieLens dataset website \url{http://files.grouplens.org/datasets/movielens/ml-100k/}. These files represent an 80\%/20\% split of the dataset into training and testing parts, respectively. Here we take $l(x,m) = - (x-m)^2/2 $ and compare the OSH method's performance with other methods as detailed in Simulation \ref{sec:simulation}. The predicted ratings are determined using the following rule:
\[
\bm{M}_{\text{pre}}= \max(1, \min(5, \hat{\bm{M}})),
\]
which corresponds to the truncation of the estimated low-rank matrix to ensure that predicted ratings fall within a plausible range.

For the comparative method, MWC \cite{MCL21}, we utilize the Akaike Information Criterion (AIC) to determine the tuning parameter for the nuclear norm penalized estimation of $\bm{\Theta}$, as illustrated in \cite{MCL21}. When estimating $\hat{\bm{M}}$ for these methods, we select the tuning parameter that yields the best mean squared prediction error (MSPE) for each split of the data. The MSPE is defined as a measure of estimation accuracy:
\[
\text{MSPE}= \frac{\sum_{(i,j) \in \text{test set}}(x_{ij}- (\bm{M}_{\text{pre}})_{ij})^{2}}{\text{cardinality
of test set}}.
\]

Additionally, to demonstrate the ranking performance of methods, we employ the modified expected percentile ranking on the test set, adapted from the approach proposed by \cite{CFA08}:
\[
\overline{\text{rank}}= \frac{\sum_{(i,j) \in \text{test set}}r_{ij}^{t}
\times \text{rank}_{ij}}{\sum_{(i,j) \in \text{test set}}r_{ij}^{t}},
\]
where $r_{ij}^{t}$ is the rating given by user $i$ to movie $j$ in the test dataset, and $\text{rank}_{ij}$ is the predicted percentile ranking of movie $j$ among all movies rated by user $i$ in the test dataset. A smaller value of $\overline{\text{rank}}$ is more desirable, with random predictions expected to yield a value of 0.5.

The OSH method consistently estimates six shared factors and no specific factors between matrices $\bm{M}$ and $\bm{\Theta}$, which corroborates our analysis of the correlation within their row spaces above. As illustrated in Table \ref{tab:realdata}, our proposed method consistently achieves the lowest MSPE, with the OSH estimator slightly outperforming the SH estimator. These results indicate that our method surpasses other methods by up to 24.02\%, underscoring the effectiveness of our approach in capturing the data's underlying structure and providing more precise predictions.

\begin{table}[htbp]
\caption{MSPE on test set and estimated ranks of $\hat{\bm{M}}$ for MovieLens
100k dataset among methods.}
\label{tab:realdata}
\begin{adjustbox}{width = 400pt}
\begin{tabular}{cccccccccccc}
\hline
\multirow{2}{*}{Method} & \multicolumn{2}{c}{Split 1} & \multicolumn{2}{c}{Split 2} & \multicolumn{2}{c}{Split 3} & \multicolumn{2}{c}{Split 4} & \multicolumn{2}{c}{Split 5} & \multicolumn{1}{c}{Overall} \\
& MSPE    & Rank    & MSPE    & Rank    & MSPE    & Rank   & MSPE   & Rank & MSPE   & Rank & MSPE   \\
\hline
OSH & \textbf{0.8906}  & 6       & \textbf{0.8609}  & 6       & \textbf{0.8604}  & 6      & \textbf{0.8657} & 6    & \textbf{0.8712} & 6    & \textbf{0.8698} \\
SH  & 0.8945  & 6       & 0.8644  & 6       & 0.8636  & 6      & 0.8688 & 6    & 0.8745 & 6    & 0.8732 \\
MHT & 1.1952  & 55      & 1.1172  & 45      & 1.1024  & 57     & 1.1226 & 56   & 1.1865 & 47   & 1.1448 \\
MWC & 1.4023  & 18      & 1.3328  & 15      & 1.3358  & 16     & 1.3368 & 16   & 1.4008 & 17   & 1.3617 \\
NW  & 1.4370  & 21      & 1.3705  & 24      & 1.3791  & 29     & 1.3882 & 21   & 1.4088 & 22   & 1.3967 \\
\hline
\end{tabular}
\end{adjustbox}
\end{table}

Furthermore, as shown in Table \ref{tab:realdata2}, our method consistently exhibits the lowest $\overline{\text{rank}}$ value across all split test data sets. This signifies that, in comparison to other methods, our method can offer more accurate rankings, which is crucial for the recommendation system's performance.

\begin{table}[htbp]
\caption{The $\overline{\rank}$ value on test set for the MovieLens
100k dataset among methods.}
\label{tab:realdata2}
\begin{adjustbox}{center}
\begin{tabular}{ccccccc}
\hline
Method     & Split 1 & Split 2 & Split 3 & Split 4 & Split 5 & Overall \\
\hline
real order & 0.4112  & 0.4085  & 0.4045  & 0.4027  & 0.4027  & 0.4059  \\
OSH        & \textbf{0.4476}  & \textbf{0.4446}  & \textbf{0.4407}  & \textbf{0.4393}  & \textbf{0.4388}  & \textbf{0.4422}  \\
SH         & 0.4477  & 0.4447  & 0.4409  & 0.4394  & 0.4388  & 0.4423  \\
MHT        & 0.4494  & 0.4461  & 0.4421  & 0.4403  & 0.4411  & 0.4438  \\
MWC        & 0.4521  & 0.4490  & 0.4454  & 0.4433  & 0.4432  & 0.4466  \\
NW         & 0.4526  & 0.4491  & 0.4455  & 0.4436  & 0.4438  & 0.4469  \\
\hline
\end{tabular}
\end{adjustbox}
\end{table}

\section{Conclusion}
\label{sec:conclusion}

In this paper, we concentrate on the matrix completion problem and introduce a shared factor structure. We establish the nonasymptotic and asymptotic statistical properties of the maximum pseudo-log-likelihood estimator under various error assumptions. Our work extends existing results in the matrix estimation problem and demonstrates the theoretical AMSE enhancement of our proposed method. To address the rank estimation challenge, we employ a matrix MCP regularization algorithm to estimate the parameter matrix, establishing a framework with theoretical guarantees, such as the Oracle property.

There are several possible subjects for future research. Firstly, we can delve into the more delicate structure between the target matrix and the missing probability parameter matrix. For instance, if both column and row spaces are partially shared, this could lead to more precise estimation techniques. Secondly, we can incorporate auxiliary side information about users and movies, as discussed in \cite{MCC18}. This information can influence both the missing probability and the rating values, and when combined with the latent factor structure, we can elucidate the relationship between the latent parameter matrices. Lastly, we can extend the shared factor model to encompass multiple data matrices, which not only include the $\bm{X}$ and $\bm{W}$ matrices discussed in this paper but could also incorporate additional matrices such as the user-item comments matrix and the user-item clicks matrix. This comprehensive approach could provide a more elaborate view of the data and enhance our understanding of user interactions and preferences.

We hope that this paper will stimulate researchers' interest in the interplay between different parameter matrices and the utilization of matrix nonconvex penalties, paving the way for more accurate estimations in the matrix completion problem.

\appendix

\section{Notation for Asymptotic Normality}
\label{sec:notation}

We denote:
\begin{align*}
\bm{\Phi}_{\theta,i} =     & \frac{1}{n_2} \sum_{j=1}^{n_2} \frac{1}{(1 + \exp(\theta_{\star, ij}))(1 + \exp(-\theta_{\star, ij}))}\begin{pmatrix}
\bm{f}_{\star,s,j} \\
\bm{f}_{\star,\theta,j}
\end{pmatrix} \begin{pmatrix}
\bm{f}_{\star,s,j} \\
\bm{f}_{\star,\theta,j}
\end{pmatrix}^\top,\\
\bm{\Phi}_{m,i} =          & -\frac{1}{n_2} \sum_{j=1}^{n_2} \pi_{\star, ij} \mathbb{E}[l^{\prime\prime}(x_{ij},m_{\star,ij})]\begin{pmatrix}
\bm{f}_{\star,s,j} \\
\bm{f}_{\star,m,j}
\end{pmatrix} \begin{pmatrix}
\bm{f}_{\star,s,j} \\
\bm{f}_{\star,m,j}
\end{pmatrix}^\top,\\
\tilde{\bm{\Phi}}_{m,i} =  & \frac{1}{n_2} \sum_{j=1}^{n_2} \pi_{\star, ij} \mathbb{E}[|l^{\prime}(x_{ij},m_{\star,ij})|^2]\begin{pmatrix}
\bm{f}_{\star,s,j} \\
\bm{f}_{\star,m,j}
\end{pmatrix} \begin{pmatrix}
\bm{f}_{\star,s,j} \\
\bm{f}_{\star,m,j}\end{pmatrix}^\top,\\
\bm{\Psi}_j =              & \begin{pmatrix}
{\bm{\psi}}_{11,j}      & {\bm{\psi}}_{12,j} & {\bm{\psi}}_{13,j} \\
{\bm{\psi}}_{12,j}^\top & {\bm{\psi}}_{22,j} & 0           \\
{\bm{\psi}}_{13,j}^\top & 0           & {\bm{\psi}}_{33,j}
\end{pmatrix}, \\
{\bm{\psi}}_{11,j} =         & \frac{1}{n_1} \sum_{i=1}^{n_1} \pi_{\star,ij}(1 - \pi_{\star,ij})\bm{\lambda}_{\star,\theta 1, i} \bm{\lambda}_{\star,\theta 1, i}^\top - \eta \pi_{\star, ij}\mathbb{E}[l^{\prime\prime}(x_{ij},m_{\star,ij})]\bm{\lambda}_{\star, m1, i} \bm{\lambda}_{\star, m1, i}^\top , \\
{\bm{\psi}}_{12,j} =         & \frac{1}{n_1} \sum_{i=1}^{n_1} \pi_{\star,ij}(1 - \pi_{\star,ij})\bm{\lambda}_{\star,\theta 1, i} \bm{\lambda}_{\star,\theta 2, i}^\top, \\
{\bm{\psi}}_{13,j} =         & -\frac{\eta}{n_1} \sum_{i=1}^{n_1} \pi_{\star,ij} \mathbb{E}[l^{\prime\prime}(x_{ij},m_{\star,ij})]\bm{\lambda}_{\star, m1, i} \bm{\lambda}_{\star, m2, i}^\top,\\
{\bm{\psi}}_{22,j} =         & \frac{1}{n_1} \sum_{i=1}^{n_1} \pi_{\star,ij}(1 - \pi_{\star,ij})\bm{\lambda}_{\star,\theta 2, i} \bm{\lambda}_{\star,\theta 2, i}^\top,  \\
{\bm{\psi}}_{33,j} =         & -\frac{\eta}{n_1} \sum_{i=1}^{n_1} \pi_{\star,ij} \mathbb{E}[l^{\prime\prime}(x_{ij},m_{\star,ij})]\bm{\lambda}_{\star, m2, i} \bm{\lambda}_{\star, m2, i}^\top,\\
\tilde{\bm{\Psi}}_j =      & \begin{pmatrix}
\tilde{\bm{\psi}}_{11,j}      & {\bm{\psi}}_{12,j} & \tilde{\bm{\psi}}_{13,j} \\
{\bm{\psi}}_{12,j}^\top         & {\bm{\psi}}_{22,j} & 0                   \\
\tilde{\bm{\psi}}_{13,j}^\top & 0           & \tilde{\bm{\psi}}_{33,j}
\end{pmatrix}, \\
\tilde{\bm{\psi}}_{11,j} = & \frac{1}{n_1} \sum_{i=1}^{n_1} \pi_{\star,ij}(1 - \pi_{\star,ij})\bm{\lambda}_{\star,\theta 1, i} \bm{\lambda}_{\star,\theta 1, i}^\top + \eta^2 \pi_{\star, ij}\mathbb{E}[|l^{\prime}(x_{ij},m_{\star,ij})|^2]\bm{\lambda}_{\star, m1, i} \bm{\lambda}_{\star, m1, i}^\top , \\
\tilde{\bm{\psi}}_{13,j} = & \frac{\eta^2}{n_1} \sum_{i=1}^{n_1} \pi_{\star,ij} \mathbb{E}[|l^{\prime}(x_{ij},m_{\star,ij})|^2]\bm{\lambda}_{\star, m1, i} \bm{\lambda}_{\star, m2, i}^\top, \\
\tilde{\bm{\psi}}_{33,j} = & \frac{\eta^2}{n_1} \sum_{i=1}^{n_1} \pi_{\star,ij} \mathbb{E}[|l^{\prime}(x_{ij},m_{\star,ij})|^2]\bm{\lambda}_{\star, m2, i} \bm{\lambda}_{\star, m2, i}^\top.
\end{align*}

\section{Correction Estimator of Variance}
\label{sec:debiasing}

For the regression case, we take the correction estimator $\hat{\sigma}_{co}^2 $ as:
\begin{equation}
\label{eq:biascorrection}
\hat{\sigma}^2_{co} = \frac{\sum_{ij}w_{ij}(x_{ij} - \hat{m}_{ij})^2 + n_2\{\tr (\bm{\Upsilon}^2) - \tr(\bm{\Upsilon})\}/\eta}{\sum_{ij} w_{ij} - n_1(d_{s} + d_m) - 2n_2 \tr(\bm{\Upsilon}) + n_2 \tr(\bm{\Upsilon}^2)},
\end{equation}
where the $\bm{\Upsilon} $ is:
\begin{align*}
\bm{\Upsilon} =& (\bm{\Upsilon}_\theta + \bm{\Upsilon}_m)^{-1} \bm{\Upsilon}_m, \\ 
\bm{\Upsilon}_\theta =& \bar{\pi}(1-\bar{\pi}) \begin{pmatrix}
\hat{\bm{\Lambda}}_{o,\theta 1}^\top \hat{\bm{\Lambda}}_{o,\theta 1} & \hat{\bm{\Lambda}}_{o,\theta 1}^\top \hat{\bm{\Lambda}}_{o,\theta 2} & 0\\
\star & \hat{\bm{\Lambda}}_{o,\theta 2}^\top \hat{\bm{\Lambda}}_{o,\theta 2} & 0\\
0 & 0 & 0
\end{pmatrix}, \\ 
\bm{\Upsilon}_m = & \eta \bar{\pi} \begin{pmatrix}
\hat{\bm{\Lambda}}_{o,m1}^\top \hat{\bm{\Lambda}}_{o,m1} & 0 & \hat{\bm{\Lambda}}_{o,m1}^\top \hat{\bm{\Lambda}}_{o,m2} \\
0 & 0 & 0 \\
\star & 0 & \hat{\bm{\Lambda}}_{o,m2}^\top \hat{\bm{\Lambda}}_{o,m2}
\end{pmatrix},
\end{align*}
here \(\bm{\Upsilon}_\theta\) and \(\bm{\Upsilon}_m\) are symmetric matrices, and $\bar{\pi} = \frac{\sum_{i,j} w_{ij}}{n_1 n_2}$ is the observation rate. The interested reader is directed to Supplement Section 18 
%  \ref*{supp-sec:biascorrection}  
for a detailed derivation of this correction estimator.

\begin{supplement}
\stitle{Supplement to "Leveraging Shared Factor Structures for Enhanced Matrix Completion with Nonconvex Penalty Regularization"}
\sdescription{This supplement material contains the computation algorithm, the proofs of the results in the main paper, and some additional numerical experiment results.}
\end{supplement}

\bibliography{reference.bib}
\bibliographystyle{imsart-number}
\end{document}